\documentclass[a4paper,a4paper,fleqn]{cas-sc}
\usepackage[numbers]{natbib}
\usepackage{placeins} 
\usepackage{siunitx} 
\usepackage{xtab} 
\usepackage{parnotes} 
\usepackage{caption} 
\usepackage{subcaption} 
\usepackage{rotating} 
\usepackage{todonotes}
\usepackage[commandnameprefix=ifneeded, draft]{changes}
\usepackage{breakcites}
\usepackage[capitalise]{cleveref}
\usepackage[export]{adjustbox}

\def\signed #1{{\leavevmode\unskip\nobreak\hfil\penalty50\hskip2em
  \hbox{}\nobreak\hfil(#1)%
  \parfillskip=0pt \finalhyphendemerits=0 \endgraf}}

\newsavebox\mybox

\PassOptionsToPackage{hyphens}{url}
\usepackage{hyperref}

\makeatletter
\g@addto@macro{\UrlBreaks}{\UrlOrds}
\g@addto@macro{\UrlNoBreaks}{\do\.} 
\makeatother

\usepackage[%
  newcommands     
 ,newparameters   
]{ragged2e}

\def\tsc#1{\csdef{#1}{\textsc{\lowercase{#1}}\xspace}}
\tsc{WGM}
\tsc{QE}
\tsc{EP}
\tsc{PMS}
\tsc{BEC}
\tsc{DE}

\begin{document}
\let\WriteBookmarks\relax
\def\floatpagepagefraction{1}
\def\textpagefraction{.001}
\shorttitle{A bibliometric analysis and scoping study to identify English-language perspectives on slums}
\shortauthors{Henn et~al.}

\title [mode = title]{A bibliometric analysis and scoping study to identify English-language perspectives on slums}                      
\tnotemark[1]

\tnotetext[1]{This document is a result of the research project "Uniform detection and modeling of slums to determine infrastructure needs" funded by the LOEWE Program of Hesse State Ministry for Higher Education, Research and the Arts. This work has been co-funded by the LOEWE initiative (Hesse, Germany) within the emergenCITY center.}

\author[1]{Katharina Henn}[orcid=0009-0006-0746-9241]
\ead{katharina.henn@tu-darmstadt.de}
\credit{Conceptualisation, Investigation, Visualization, Writing - First Draft, Writing - Review and Editing}

\author[1]{Michaela Le\v{s}táková}[orcid=0000-0002-5998-6754]
\ead{michaela.lestakova@tu-darmstadt.de}
\credit{Conceptualisation, Writing - Review and Editing}

\author[1]{Kevin Logan}[orcid=0000-0001-5512-2679]
\ead{kevin.logan@tu-darmstadt.de}
\credit{Conceptualisation, Writing - Review and Editing}

\author[3]{Jakob Hartig}[orcid=]
\ead{j.hartig@hartig-germany.com}
\credit{Conceptualisation, Writing - Review and Editing}

\author[2]{Stefanos Georganos}[orcid=]
\ead{stefanos.georganos@kau.se}
\credit{Writing - Review and Editing}

\author[1,4]{John Friesen}[type=editor,
                        auid=000,bioid=1,
                        orcid=0000-0002-6226-7208]

\address[1]{Chair of Fluid Systems, TU Darmstadt, Otto-Berndt-Strasse 2, 64287 Darmstadt, Germany}
\ead{john.friesen@tu-darmstadt.de}
\cormark[1]
\credit{Conceptualisation, Methodology, Investigation, Writing - First Draft, Writing - Review and Editing}

\cortext[cor1]{Corresponding author}

\address[2]{Geomatics, Department of Environmental and Life Sciences, Karlstad University, Karlstad, Sweden}
\address[3]{Hartig GmbH \& Co KG, Lohrweg 6, 63741 Aschaffenburg, Germany}
\address[4]{Earth Observation Research Cluster, University of Würzburg, John-Skilton-Str. 4a, 97074 Würzburg, Germany}

\begin{abstract}
Slums, informal settlements, and deprived areas are urban regions characterized by poverty. According to the United Nations, over one billion people reside in these areas, and this number is projected to increase. Additionally, these settlements are integral components of urban systems. We conducted a bibliometrical analysis and scoping study using the Web of Science Database to explore various perspectives on urban poverty, searching for scientific publications on the topic and providing details on the countries where the studies were conducted. Based on 3947 publications, we identify the extent to which domestic research organizations participate in studying urban poverty and which categories of science they investigate, including life sciences \& biomedicine, social sciences, technology, physical sciences, and arts \& humanities. Thereby, we find that research on slums is often limited to specific countries, e.g. India, South Africa, Kenya, or Brazil. This focus is not necessarily correlated with the number of people living in slums. The scientific discourse is up to now predominantly shaped by medical and social sciences with few studies addressing technological questions. Finally, our analysis identifies several possible future directions for research on slums.
\end{abstract}

\begin{graphicalabstract}
\includegraphics[width=\linewidth]{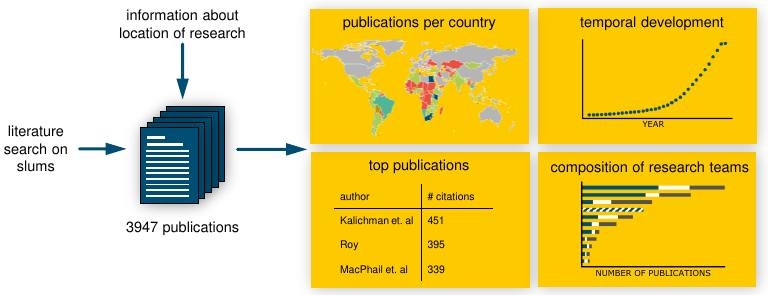}
\end{graphicalabstract}

\begin{highlights}
\item Thematic and geographical analysis of 3947 publications on slums
\item Research on slums is mainly focused on specific areas like India, South Africa, Kenya or Brazil
\item Studies from medical and social sciences shape the scientific discourse while technological studies are rare
\item Identification of potential paths for future slum research 
\end{highlights}


\maketitle

\section{Introduction}
\begin{figure}[h!]
\centering
\includegraphics[width=\textwidth]{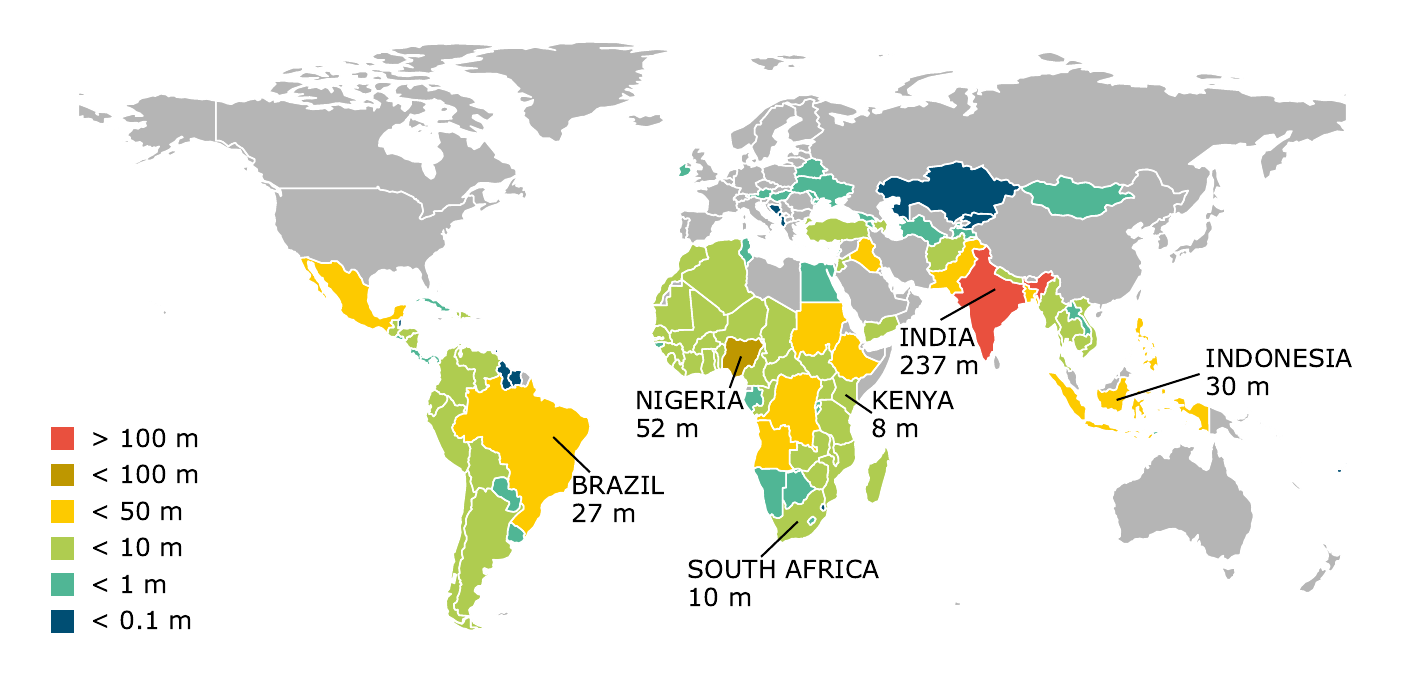}
\caption{Population living in slums in million (m) according to data from UN Habitat World Cities report \cite{UNHabitat.2022}.Gray color represents countries without available data.}
\label{fig:slumpopu}
\end{figure}

Since the Millenium Development Goals were published by the United Nations (UN), deprived urban areas, often referred to as slums, have not only become more and more present on political agendas \cite{GILBERT.2007}, but also have become the object of an increasing number of scientific publications \cite{Mahabir.2016}. Slums are defined by the UN as areas or households in which at least one of the following five criteria is met: (i) inadequate access to safe water, (ii) inadequate access to sanitation, (iii) poor structural quality of housing, (iv) overcrowding or (v) insecure residential status.  According to the UN Habitat Cities report \cite{UNHabitat.2022}, more than one billion people live in slums. These populations are distributed across the globe as shown in Figure \ref{fig:slumpopu}. Most slum residents, according to these numbers, live in India (237 Mio.), Nigeria (52 Mio.), and Indonesia (30 Mio.).

With the rapid growth of cities, which is quickly becoming an important global challenge, the fact that slums are receiving more attention is not surprising. Nevertheless, the focus, not only of scientific publications \cite{Friesen.2020}, but also of pop culture, as Roy impressively showed \cite{Roy.2011}, is often on big, well-known slums: Orangi in Karachi (Pakistan), Dharavi in Mumbai (India), Kibera in Nairobi (Kenya) or Paraisopolis in Sao Paulo (Brazil) are often refered to, although the majority of the dwellers in the respective cities seem to live in smaller, though more numerous settlements \cite{Friesen.2018, Breuer.2022}. These face the risk of remaining “invisible” in research. 

Moreover, research on slums is characterized by discourses of power, which is even evident in the negative connotation of the word slum \cite{Mayne.2017}. This term is often not used as a self-designation but imposed on the settlement areas and their inhabitants from the outside \cite{GILBERT.2007}. Slum dwellers are often the object of analysis and rarely enter the research discourse as participating subjects. These problems have already been raised and discussed thoroughly \cite{GILBERT.2007, Mayne.2017, Roy.2011}. In order to obtain a more comprehensive perspective on urban poverty, the \textit{Domains of deprivation framework} has therefore been developed in recent years \cite{Abascal.2022}. It seeks a more comprehensive understanding of deprivation and also refers to settlements as \textit{deprived areas} rather than slums. Although the authors agree, that the term \textit{deprived areas} addresses the question of disadvantages in urban areas more comprehensively, we will continue to use the term slums in the following, as it is used within the Millennium and Sustainable Development Goals and thus has dominated research in recent decades.

Several studies have examined different aspects of slums: Ezeh et al.~\cite{Ezeh.2017} summarised the state of the art on the health of slum dwellers, noting that the vulnerability of children is one of the main issues in the scientific literature on health and slums. However, there are still many unanswered questions about health in these settlements. It has also been shown that many studies on slums and health have been conducted in very specific countries and cities (such as Kenya, South Africa or India) and that some countries with large numbers of people living in slums are under-represented \cite{Friesen.2020}. 

Another category of reviews (e.g. \cite{Kuffer.2016, Mahabir.2016}) deals with the classification, identification and mapping of slums using modern technologies such as remote sensing. Abascal et al.~\cite{Abascal.2022} extended and updated this work on slum mapping in a scoping study, showing that there are no scalable methods for mapping slums in low and middle income countries. Similar to the medical studies mentioned above, these reviews showed that knowledge on slums is often concentrated in a few countries.

Other reviews and conceptual works focus on the population distribution~\cite{Thomson.2021, Breuer.2022} in slums, their economics~\cite{Marx.2013} or their importance for urban transformations from an epistomological point of view \cite{roy.2005}.  

To date, there has been no overview of the state of slum studies that takes into account the different disciplines and shows the development of slum studies over time. There are also no known works that identify frequently discussed topics in the context of slums for which the state of research has been summarised in the form of reviews. 

In addition, a number of publications in recent years have pointed out that perspectives in scholarly publications are not evenly distributed in terms of race, ethnicity \cite{Hrabowski.2020} or gender \cite{Huang.2020, Rosenman.2020}, which in turn has an impact on the focus of research. In the context of slums, it is particularly important to reflect on the researcher himself, as perspectives on slums in the past have often been characterised by (negatively connoted) attributions and descriptions from the global North, as Mayne \cite{Mayne.2017} or Gilbert \cite{GILBERT.2007} impressively demonstrate.

Under these circumstances, it is high time/urgently needed to reflect on the past and current pathway of research on slums and to map our state of knowledge. Combining a bibliometrical analysis \cite{Biswas.2021} with a scoping review, we aim to answer the following research questions:
\begin{enumerate}[(i)]
    \item \textbf{Geographic location of the investigated slums:} Which countries and regions were primarily the object of research in the study period? Is the research representative with regard to the slum population? 
    \item \textbf{Geographic location of the researchers investigating slums:} With which organizations are the researchers affiliated and where are these organizations based?  How is the knowledge of slums spatially distributed?
    \item \textbf{Investigated topics/themes:} Which topics/themes stand in the foreground?

\end{enumerate}

In other words, we aim to investigate “who (researching country) is researching whom (researched country) and what (research theme/topic)” and relate the results to the relative slum population to obtain a metric for identifying blind spots. We also aim to investigate the publications' development in time.

\section{Methods}
\label{sec:Material+Methods}
To answer the research questions, we combine two types of literature analysis: a bibliometric study and a scoping study. While in a bibliometric study the literature is assessed in a quantitative way \cite{andres2009measuring} by analysing number of publication, author institutions or citations, a scoping study aims to "rapidly map the key concepts underpinning a research area and
the main sources and types of evidence available" \cite{arksey2005scoping}. Both types of studies help to describe the current state of of the art with the aim to identify gaps in the body knowledge \cite{arksey2005scoping}. Thereby, it is noteworthy that theses methodologies do not identify gaps in the research due to poor quality, since the quality assessment of the studies is not part of bibliometrical analyses or scoping studies \cite{andres2009measuring, arksey2005scoping}. 

The basic stages of a scoping review are: (i) identifying the research question, (ii) identifying relevant studies, (iii) study selection, (iv) charting the data and (v) collating, summarising and reporting the results.

Since the research questions (stage (i)) are described above, the next stage (ii) is identifying relevant studies.
The publications our analyses are based on were identified through a search in the Web-Of-Sciene (WoS) Database. We use WoS due to its high amount of available meta data in comparison to other databases \cite{Kokol.2018}. The search terms used are slum, informal settlement, as general descriptions of areas of urban poverty enriched by specific local names like barrio, bidonville or favela. We only searched for publications with the search term within the title of the study. Thereby, we wanted to ensure, that this type of settlements is in the core of the paper and not a side aspect, mentioned in the abstract. The comprehensive query can be found in Sec.~\ref{sec:searchmethodology} and is based on the search methodology documented in Ezeh et al. \cite{Ezeh.2017}.

The search was performed on March 4, 2022. Therefore, our database and analysis is limited to publications that were published before these dates. In addition, the metadata of the identified publications, such as the number of citations, may not correspond to the current value of this figure. Only English-language publications are included in our search, as it would otherwise be more difficult to assess whether the publication matches our search terms. 

In the third stage of a scoping study relevant studies are selected: Therefore, two of the authors initially selected publications with available abstract and subsequently screen all abstracts to ensure their relevance to research on urban slums. Due to this methodology, articles from scientific disciplines publishing primarly in books without abstracts will not appear in our final database. Using this methodology, we created a unique database, available in the Supplementary Materials. This database forms the foundation of the following analyses and can be used by other researchers as a starting point for additional studies.

The data extracted from the WoS database includes the metadata of the publications, such as the number of citations or the authors' affiliation. Some of the metadata of the 3947 publications analysed are incomplete. For 4\,\% of the publications the year of publication is missing and for 3\,\% of the publications no affiliation is assigned to the authors. Nevertheless, publications with missing data are included in the analysis as far as possible.

During our analysis, the WoS database is enriched by using some publication attributes. 

Based on the created data base with the additional data, we charted the literature using different techniques from bibliometrical analyses, like the categorization according countries, time or thematic focus of the respective studies. 

Finally, we summarized the findings and identified possible directions for further research. The process of enriching the database with additional data is described below. 

\subsection*{Researched Countries}
To be able to use the geographical location of each study for our analysis, we use the manual screening of titles and abstracts to assign a publication to the country the study was performed in, this attribute is hereinafter referred to as the researched country of a publication. If a publication addressed research in more than one country the geographic area of research is specified as \textit{international} in our database. If no geographic area was apparent from the abstract of a publication, the geographic area was classified as \textit{international} as well. Examples for this category are general reviews without a specific geographical focus like \cite{Kuffer.2016,Mahabir.2016}. 

In section ~\ref{sec:limitation} we discuss the shortcomings resulting from the manual assignment to research countries, as they significantly influence our results.

\subsection*{Researching Countries}
To gain information on the geographical context of a publications authors, we introduced the attribute \textit{researching country}. Most of the publications have in the address section information on the affiliations of the authors. We use these affiliations to assign each publication information on the \textit{researching country}. Since some publications are the result of a cooperation between authors from different institutions, one publication can have multiple researching countries.
Based on this information we also distinguish between different types of researching groups based on the researching countries:
\begin{itemize}
        \item \textit{domestic}: All researching countries of the publication are identical with the researched country studied in this publication.
        \item \textit{foreign}: All researching countries differ from the researched country.
        \item \textit{collaborative}: At least one researching country is different from the researched country and at least one researching country is identical to the researched country .
\end{itemize}

\subsection*{World Regions}
In addition, the researched countries were grouped into georaphic world regions based on the definition of the World Bank \footnote{\url{https://datahelpdesk.worldbank.org/knowledgebase/articles/906519-world-bank-country-and-lending-groups}}. The categorization into the world regions \textit{East Asia \& Pacific}, \textit{Europe \& Central Asia}, \textit{Middle East \& North Africa}, \textit{North America}, \textit{Latin America \& the Caribbean}, \textit{South Asia}, and \textit{Sub-Saharan Africa} provides a high-level overview on the geographic focus of research on slums.

\subsection*{Population Living in Slums}
To allow a comparison between the number of publications on slums in a given country (e.g. number of publications with the attribute researched country equals the country) and the number of people living in slums in that country, we used data from the UN Habitat World Cities Report \cite[p.348 f.]{UNHabitat.2022}. We have used the most recent data where available (2020), which gives the number of people living in slums at country level. These figures are linked to the number of publications per country analysed in this study. Note that only publications with a researched country not equal to \textit{international} and countries with data on number of slum dwellers can be considered when creating the ratio described above.

\subsection*{Categories \& Research Areas}
The metadata of publications extracted from the WoS database contain \textit{WoS categories}. The assignment to a \textit{WoS category} is based on the journal or book the publications are published in, as \textit{WoS categories} are assigned on the source publication level \footnote{See \url{https://support.clarivate.com/ScientificandAcademicResearch/s/article/Web-of-Science-Core-Collection-Web-of-Science-Categories?language=en_US} and \url{https://images.webofknowledge.com/images/help/WOS/hp_subject_category_terms_tasca.html}}. Therefore, it should be noted that the categories reflect the topics covered in the journals or books in which the publication appeared rather than the topic of the publications themselves. A book or journal can be assigned to up to 6 \textit{WoS categories} \footnote{See \url{https://support.clarivate.com/ScientificandAcademicResearch/s/article/Web-of-Science-Core-Collection-Web-of-Science-Categories?language=en_US}}. One publication in our dataset is not assigned to a  \textit{WoS category} and is therefore excluded from the analysis in Sec.~\ref{sec:Categories}. For the sake of clarity, we have merged thematically similar WoS categories. Such categories are marked with * in Sec.~\ref{sec:Categories}. One example is "Agriculture and Forestry*", which combines "Agricultural Economics and Policy", "Agriculture, Multidisciplinary", "Agronomy", and "Forestry". Otherwise we used the categories as defined by the WoS.

Since the data set studied here comprises 196 different \textit{WoS categories}, the categories were grouped into superordinate  \textit{research areas}, where possible in accordance with \footnote{\url{https://images.webofknowledge.com/images/help/WOS/hp_research_areas_easca.html}}. If a publication's \textit{WoS categories} belonged to two or more different research areas, the thematic focus of the publication is considered to be \textit{Multidisciplinary Sciences}. Besides the we used the research areas \textit{Life Sciences \& Biomedicine}, \textit{Social Sciences}, \textit{Technology}, \textit{Physical Sciences} and \textit{Arts \& Humanities}, based on the \textit{WoS categories}, to assess the thematic focus of research on urban poverty.

\subsection*{Keywords}
Part of the WoS metadata of a publication are the \textit{Author Keywords}, more than 80\,\% of the publications in our database have this attribute. We use the keywords to get an insight into the topic of publications.

\subsection*{Citations}
When determining publications in our database that have a high impact, we have taken into account the attribute \textit{Times Cited, All Databases}. This number may differ from the number reported by other databases (e.g. Google Scholar).

For access to highly relevant research institutions (research institutions that are frequently included in a publication affiliation) and journals, we calculated the h-index, which is based on the number of citations of a publication in all databases. Since it is not possible to calculate the h-index of a single publication, we had to limit ourselves to the number of citations when determining popular publications.

\subsection*{Publication Year}
For a temporal analysis of the publications, we considered the WoS attribute \textit{year of publication}. More than 96\,\% of the publications analysed have this attribute and are included in this part of our study.

\section{Results}

\begin{figure}[h!]
\centering
\includegraphics[width=0.9\textwidth]{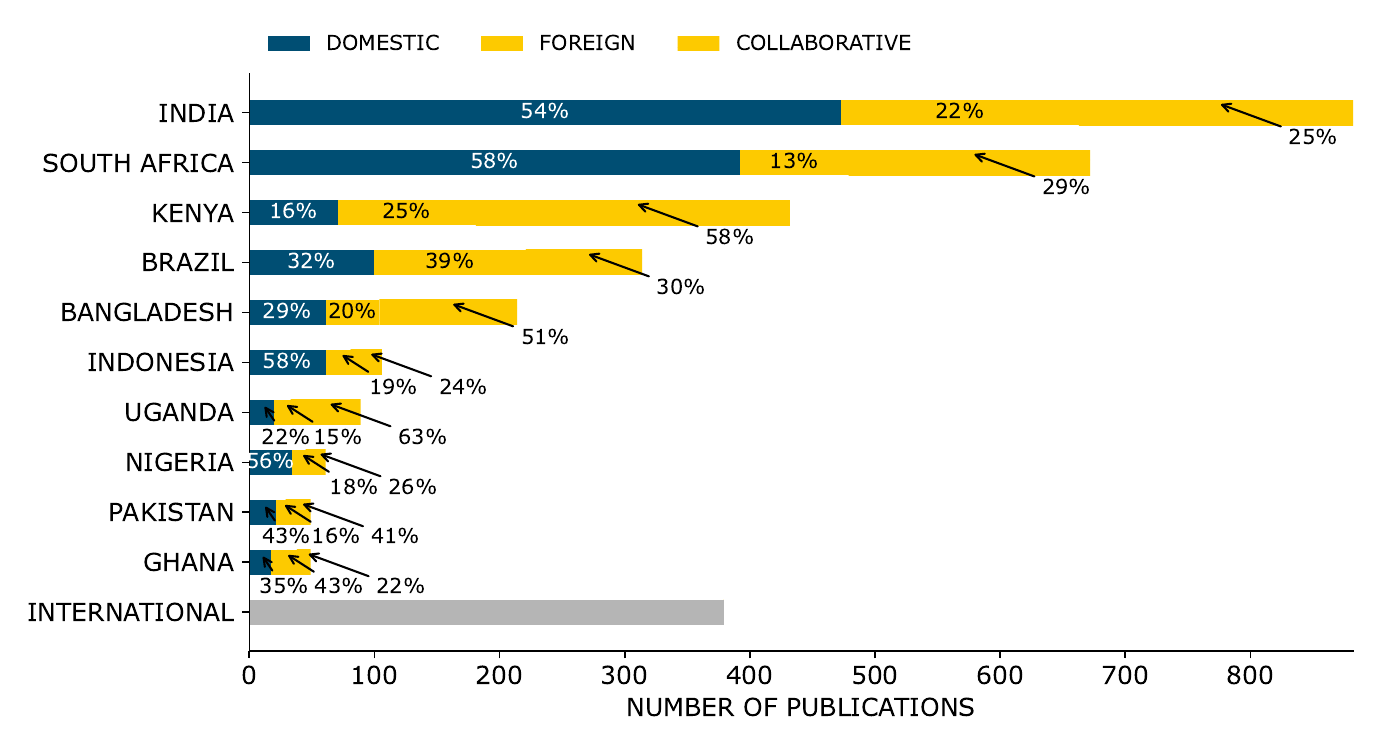}
\caption{Publications assigned to the ten most frequently researched countries and to the category \textit{international}. The data is divided into three different types by origin of research institutions.}
\label{fig:reasearched_bar}
\end{figure}

The number of publications for the ten most researched countries in our literature search (and the category international) are shown in
Figure~\ref{fig:reasearched_bar}, representing $84$\,\% of the total data set. 

Besides (i) depicting the countries where the most research is conducted on urban slums, Figure~\ref{fig:reasearched_bar} also shows (ii) the geographic composition of the research institution(s) of the publications for the different researched countries. The temporal development of the geographic composition of research institution(s) is shown in Figure~\ref{fig:mixed_foreign_native_over_years}.

India is the country most frequently studied, followed by South Africa, Kenya, Brazil and Bangladesh. 
This is indeed influenced by the terminology used to describe slums. In the literature with researched countries Brazil and South Africa, for example, the terms "favela" and "township" are frequently used, but their definitions are often broader than "urban informal settlement" or "slum". This could imply that South Africa and Brazil are over represented in the data set from Figure~\ref{fig:reasearched_bar} as the terms are applied to more types of urban living than just slums.

\begin{figure}[h!]
\centering
\includegraphics[width=\textwidth]{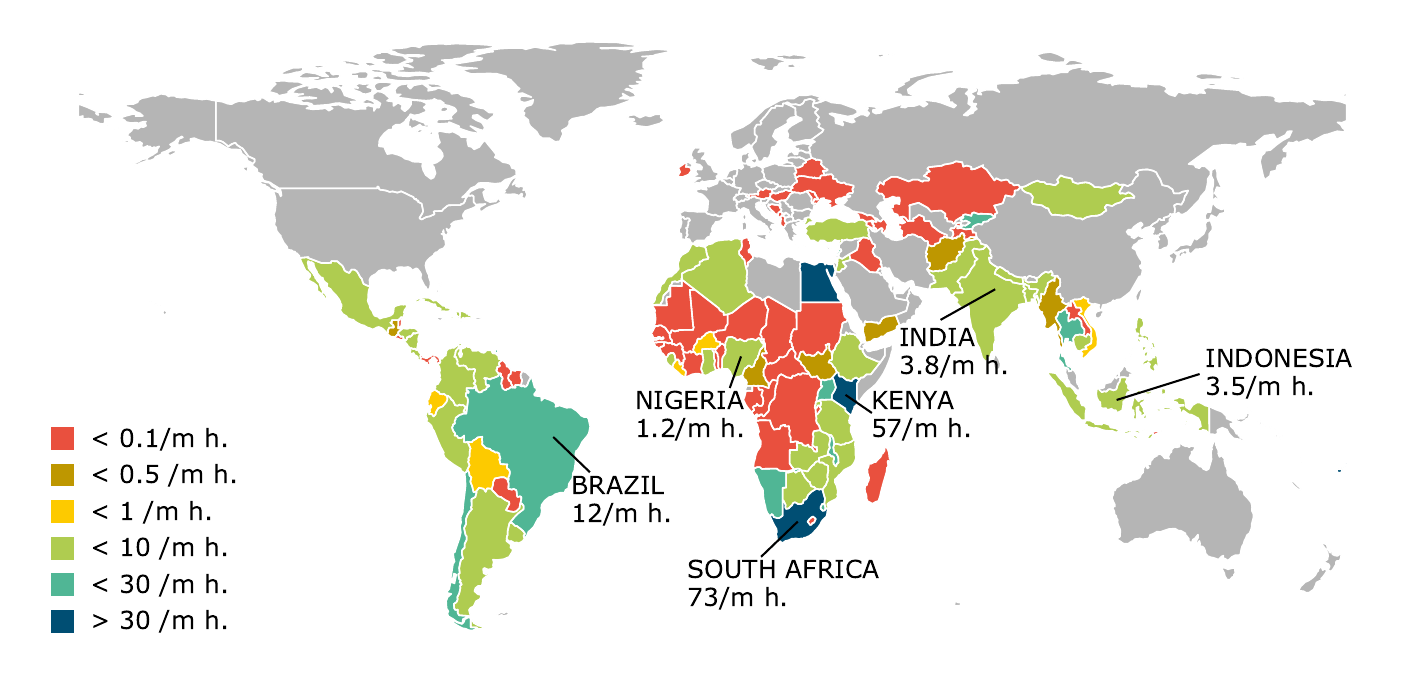}
\caption{Number of publications about a country in relation to the million inhabitants (m h.) of slums.}
\label{fig:studperinh}
\end{figure}

It is noteworthy that research on slums in Nigeria is represented by a total of 61 publications, representing 1.54\,\% of the data set studied. Yet, residents of slums in Nigeria represent 4.95\,\% of the total population living in slums worldwide, according to data from \cite{UNHabitat.2022}. This can also be seen in Figure~\ref{fig:studperinh}, where the number of studies assigned to each country is set in relation to the number of people living in slums in that country. This serves as a visibility metric that can be used to highlight the relatively well-researched areas and relatively less-researched areas. Another representation of this data is available in the Appendix in Figure~\ref{fig:ScatterSlumPopVsNationalStudies}.

While India, a certain number of South American countries, and some East African countries (South Africa, Kenya, Namibia, Zimbabwe) show a high amount of studies per population, urban poverty in West Africa seems to be severely lacking in the literature. Slums and settlements of urban poverty in Central Africa and in the states east of India are also strongly underrepresented in the scientific literature. 

With regard to aspect (ii), it can be said that in general, most groups of institutions involved in a publication are of the collaborative or domestic type. Research countries that deviate from this are Brazil, where a share of 39\,\% of the publications is written without a contribution from a Brazilian institution. The opposite can be observed for Kenya, where only 16\,\% of all publications are written exclusively by researchers affiliated with institutions from Kenya.

\begin{figure}[h!]
\begin{subfigure}[b]{0.5\textwidth}
\centering
\includegraphics[width=1\textwidth]{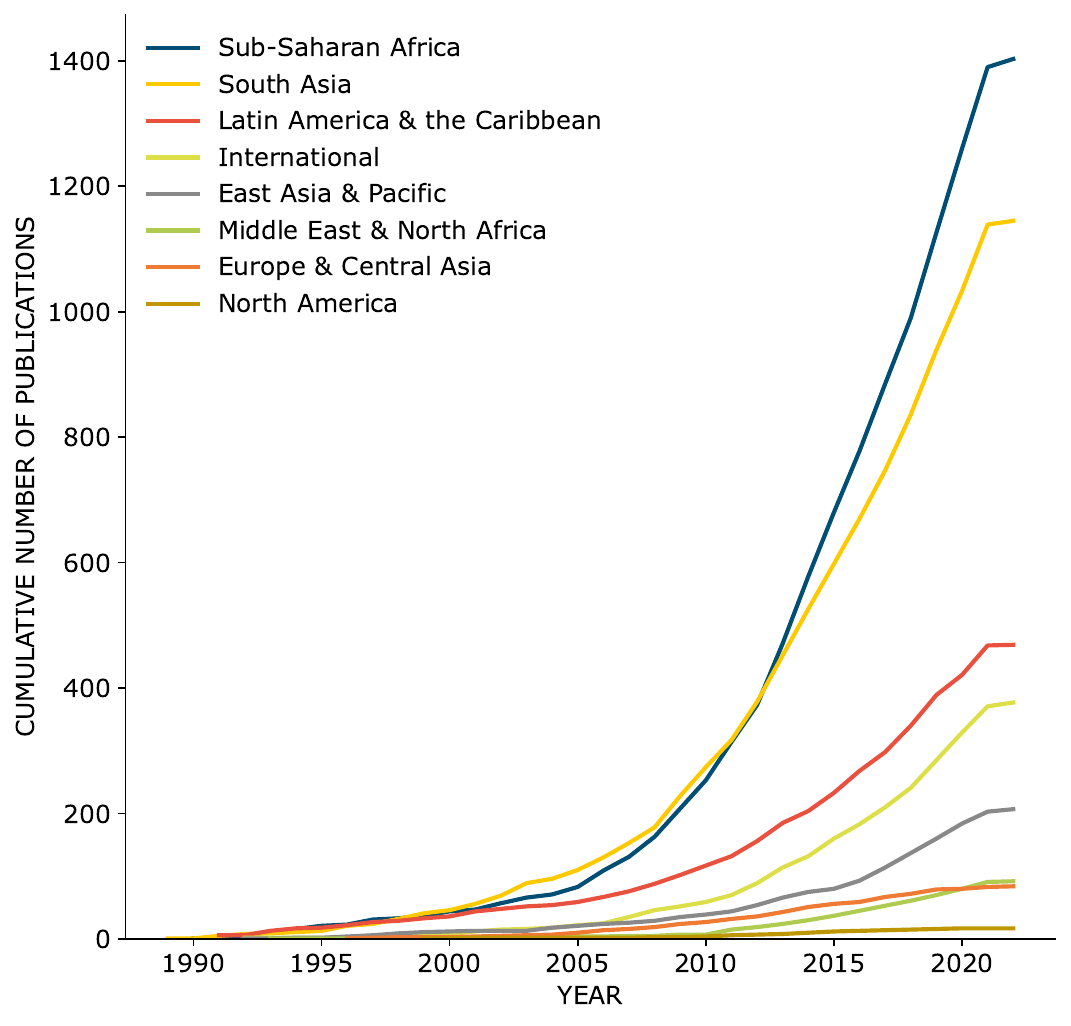}
\caption{Cumulative publications grouped by world region}
\label{fig:temporalWorldRegion}
\end{subfigure}
\begin{subfigure}[b]{0.5\textwidth}
\centering
\includegraphics[width=1\textwidth]{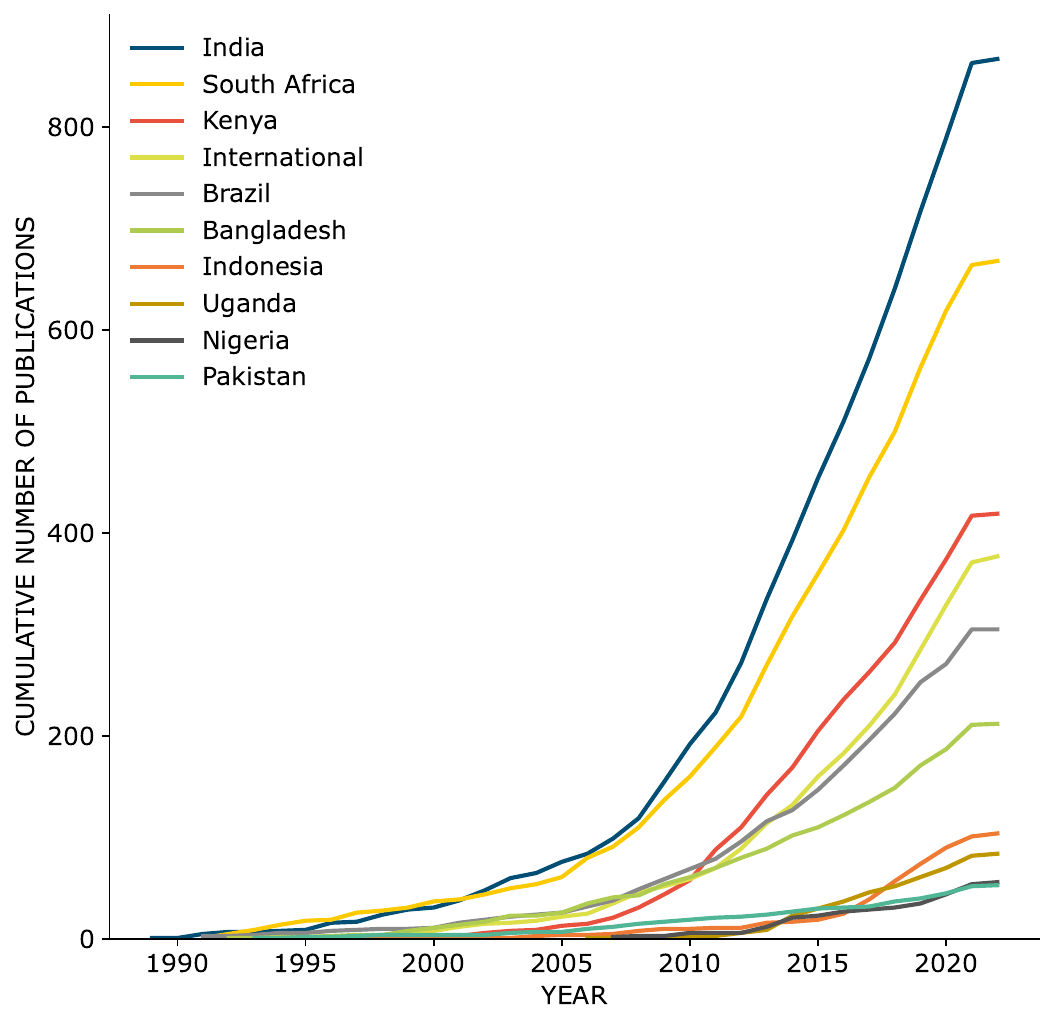}
\caption{Cumulative publications per top 10 researched country}
\label{fig:temporalCountryResearched}
\end{subfigure}
\caption{Time development of the number of publications by researched country.}
\end{figure}

Figure~\ref{fig:temporalWorldRegion} and Figure~\ref{fig:temporalCountryResearched} show the temporal development of publications addressing slums categorized by world regions and by researched countries. The total number of publications grows exponentially with a growth rate of 10.9\,\% per year. Compared to growth rate in the same time period given by Bormann and Mutz \cite{Bornmann.2015}, the growth rates are higher than in science in general. The values used for comparison can be found in Table~\ref{tab:Growthrates} in the Appendix.

If we look at the temporal development of publications on a country-level (Figure \ref{fig:temporalCountryResearched}), the strong increase in research on slums could be a correlation with the adoption of the Millennium Development Goals of the UN in 2000. This in turn builds on the Cities Without Slums initiative of the Cities Alliance in 1999 \cite{GILBERT.2007}.
While the number of publications are similar for Sub-Saharan Africa and South Asia until 2012, Sub-Saharan Africa shows a higher cumulative number in recent years (see Figure~\ref{fig:temporalWorldRegion}) .
The rising number of publications in Kenya since 2005 could be connected to the often cited Nairobi Urban Health and Demographic Surveillance System (NUHDSS) program, which several of the publications refer to \cite{Kyobutungi.2008}.

\subsection{Most influential publications}
\label{sec:MostInfluentialPublications}
\begin{sidewaystable}
\small
\def\arraystretch{2.5}
    \caption{The ten most cited publications in the data set. The column "Times Cited" corresponds to the column "Times Cited, All Databases" in WoS metadata, "Year" corresponds to "Publication Year". WoS categories are separated by ';' and researching countries are sorted alphabetically.}
    \centering
    \resizebox{\textwidth}{!}{
    \begin{tabular}{l l l l l l l}
        
        \parbox{0.8cm}{\textbf{Times Cited}} & \textbf{Article Title} & \textbf{Authors} & \textbf{Year} & \textbf{WoS Categories} & \textbf{Researched Country} & \parbox{2.7cm}{\textbf{Researching Countries}} \\  [0.2cm]  \hline 
        451 & \parbox{6cm}{HIV testing attitudes, AIDS stigma, and voluntary HIV counselling and testing in a black township in Cape Town, South Africa} \cite{Kalichman.2003} & \parbox{4cm}{Kalichman, SC; Simbayi, LC } & 2003 & Infectious Diseases & South Africa & South Africa, USA
        \\ [0.8cm]  \hline
        
        395 & \parbox{6cm}{Slumdog Cities: Rethinking Subaltern Urbanism} \cite{Roy.2011} & Roy, A & 2011 & \parbox{4cm}{Geography; Regional \& Urban Planning; Urban Studies} & International & USA
        \\ [0.8cm] \hline
        
        339 & \parbox{6cm}{'I think condoms are good but, aai, I hate those things': condom use among adolescents and young people in a Southern African township} \cite{MacPhail.2001} & \parbox{4cm}{MacPhail, C; Campbell, C}  & 2001 & \parbox{4cm}{Public, Environmental \& Occupational Health; Social Sciences, Biomedical} & South Africa &  South Africa, UK
       \\ [0.5cm] \hline
        
        303 & \parbox{6cm}{Exclusive breastfeeding reduces acute respiratory infection and diarrhea deaths among infants in Dhaka slums} \cite{Arifeen.2001}& \parbox{5cm}{Arifeen, S; Black, RE; Antelman, G; Baqui, A; Caulfield, L; Becker, S}  & 2001 & Pediatrics & Bangladesh & USA
        \\ [0.5cm] \hline
        
        248 & \parbox{6cm}{High prevalence of diabetes, obesity and dyslipidaemia in urban slum population in northern India} \cite{Misra.2001} & \parbox{5cm}{Misra, A; Pandey, RM; Devi, JR; Sharma, R; Vikram, NK; Khanna, N} & 2001 & \parbox{4cm}{Endocrinology \& Metabolism; Nutrition \& Dietetics} & India & India
        \\ [0.5cm] \hline

        234 & \parbox{6cm}{Effect of anthelmintic treatment on the allergic reactivity of children in a tropical slum}\cite{Lynch.1993} & \parbox{5cm}{Lynch, NR; Hagel, I; Perez, M; Di Prisco, MC; Lopez, R; Alvarez, N} & 1993 & Allergy; Immunology & Venezuela & Venezuela
        \\ [0.5cm] \hline
        
        216 & \parbox{6cm}{The state of emergency obstetric care services in Nairobi informal settlements and environs: Results from a maternity health facility survey} \cite{Ziraba.2009}& \parbox{5cm}{Ziraba, AK; Mills, S; Madise, N; Saliku, T; Fotso, JC} & 2009 & \parbox{4cm}{Health Care Sciences \& Services} & Kenya & Kenya, UK, USA
        \\ [0.5cm] \hline
        
        207 & \parbox{6cm}{Impact of Environment and Social Gradient on Leptospira Infection in Urban Slums} \cite{Reis.2008}& \parbox{5cm}{Reis, RB; Ribeiro, GS; Felzemburgh, RDM; Santana, FS; Mohr, S; Melendez, AXTO; Queiroz, A; Santos, AC; Ravines, RR; Tassinari, WS; Carvalho, MS; Reis, MG; Ko, AI} & 2008 & \parbox{4cm}{Infectious Diseases; Parasitology; Tropical Medicine} & Brazil & Brazil, USA
        \\ [0.5cm] \hline
        
        166 & \parbox{6cm}{Making the best of a bad situation: Satisfaction in the slums of Calcutta} \cite{BiswasDiener.2001} & \parbox{5cm}{Biswas-Diener, R; Diener, E} & 2001 & \parbox{4cm}{Social Sciences, Interdisciplinary; Sociology} & India  & USA  \\ [0.5cm] \hline

        164 & \parbox{6cm}{The return of the slum: Does language matter?} \cite{GILBERT.2007} & \parbox{5cm}{Gilbert, A} & 2007 & \parbox{4cm}{Geography; Regional \& Urban Planning; Urban Studies} & International & UK 
        
    \end{tabular}}
    \label{tab:TopPupblications}
\end{sidewaystable}

To explain the data set in more detail, we show the ten most cited publications in Table~\ref{tab:TopPupblications}. Information regarding the researched country as well as the researching countries are also shown. While the research team of the first study is classified as \textit{collaborative}, since the reasearching countries are South Africa and the USA, the fourth study is classified as \textit{foreign}, since the USA are the researching country while the researched country is Bangladesh. In contrast, the researching country of the fifth study is classified as \textit{domestic}, since researching and researched country are both India.

Thematically, most of the publications in Table~\ref{tab:TopPupblications} deal with health issues and questions of reproductive health in slums. Kalichman and Simbayi \cite{Kalichman.2003} analyse HIV testing attitudes and AIDS stigma, MacPhail and Campbell \cite{MacPhail.2001} analyse the use of condoms in these settlements, Arifteen et al. \cite{Arifeen.2001} analyse the relationship between breastfeeding and infections diseases in Bangladesh and Misra et al. \cite{Misra.2001} the prevalence of diabetes in Indian slums. Lynch et al. \cite{Lynch.1993} investigate allergic reactivity in Venezuela, Ziraba et al. \cite{Ziraba.2009} analyse health care services in Nairobi, Kenya and Reis et al. \cite{Reis.2008} the impact of environment on specific infections in Brazil. 

These studies reflect the thematic aspects most discussed in literature when it comes to slums. The insufficient access to different kinds of infrastructure, such as water or sanitation, healthcare or education, makes these areas vulnerable to different kind of health issues like infections or cardiovascular diseases \cite{Corburn.2012, Ezeh.2017, Friesen.2020b}. These publications also reflect the reasons why slums are explicitly mentioned in the UN sustainable development goals, since they describe explicitly the different dimensions of vulnerability within these areas \cite{Lilford.2019}. 

Looking at the three other publications in the list, another aspect of research on urban slums becomes clearer. These studies from urban and social sciences discuss the perspectives on slums from the outside (Roy \cite{Roy.2011}, \cite{GILBERT.2007}) or the perspective dwellers have of their own situation (Biwas-Diener and Diener \cite{BiswasDiener.2001}). The very famous article of Roy \cite{Roy.2011} is an example of studies categorized as \textit{international}. Although Roy builds her article around Indian slums (referring to the Oscar winning movie Slumdog Millionare), she discusses the topic with regard to the global phenomenon of slums. The article \cite{GILBERT.2007} focuses on the possible implications of the use of the term "slums" in contexts such as the "Cities without Slums" initiative and thus deals with research that we classify as \textit{international} in this analysis.

The interested reader is referred to the appendix, which contains lists of the most frequently cited publications from the different research areas in our database. There it can be observed, that the top publications in the category \textit{Technology} mainly deal with the identification and mapping of slums, while the category \textit{Physical Sciences} is focused on flooding, sanitation, disaster management or resilience. The top publications in \textit{Arts \& humanities} mainly deal with historical or cultural aspects of slums and slum tourism, while the \textit{Social Sciences} deal with property rights and residential satisfaction. 

\subsection{Categories}
\label{sec:Categories}
In this section, we analyse the categories and research areas the different publications deal with in more detail.

\begin{figure}[h!]
\centering
\includegraphics[width=0.8\textwidth]{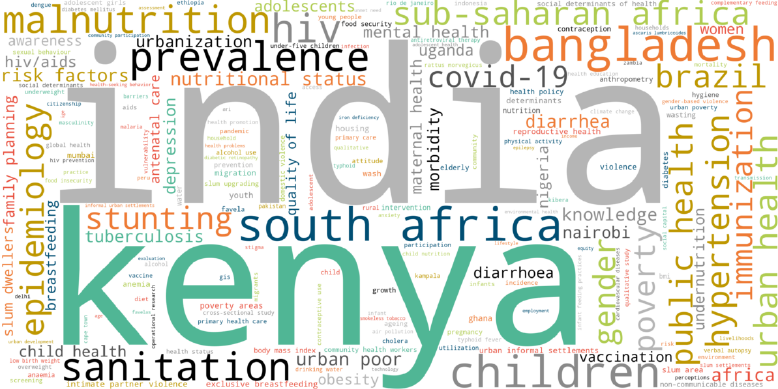}
\caption{Word cloud consisting of author keywords for publications with research area \textit{Life Sciences \& Biomedicine}. Generic and obvious keywords such as "health", "urban slum" and "informal settlement" are not taken into account.}
\label{fig:KeywordsLifeScience}
\end{figure}

To gain inside into all, and not only the most cited publations connected to the research area \textit{Life Science \& Biomedicine} the most frequently used author keywords are shown in Figure~\ref{fig:KeywordsLifeScience}. Besides showing the geographical focus on India, Kenya, Bangladesh and South Africa (which is investigated in more detail later), general thematic spotlights become apparent. Multiple keywords are connected to pregnancy and child birth such as "breastfeeding", "exclusive breastfeeding", "maternal health" and "pregnancy". Malnutrition and diet is a second thematic focus as intended by the keywords "malnutrition", "nutritional status", "undernutrition", "diet", "stunting" and "body mass index". It is also clear, that communicable diseases like "Covid-19", "diarrhea", "HIV", "tuberculosis" and connected topics like "immunization" or "vaccination" are investigated. The frequency of the most frequently used keywords in shown in Table~\ref{tab:frequencyKeywordsLifeScience}. 

When repeating the process of creating a word cloud for publications from the other research areas (Figure~\ref{fig:KeywordsSocialScience},\ref{fig:KeywordsArts}, \ref{fig:KeywordsTech}, \ref{fig:KeywordsPhysi}, \ref{fig:KeywordsMulti}), the different thematic focuses of the research areas become clear: While \textit{Life Science \& Biomedicine} has a focus on medicine (especially HIV, malnutrition, epidemiology and maternal health), \textit{Technology} discusses topics as "remote sensing" and "fire spread". While \textit{Physical Science} focuses on topics connected to water such as "flooding", "water quality", "water supply" and "sanitation".

Research in the field of \textit{Arts \& Humanities} focuses on cultural phenomena such as the movie Slumdog Millionaire and Bollywood. The two Brazilian politicians Luiz Paulo Conde and Sergio Magalheas are also addressed. The keywords on \textit{Social Sciences} indicate a broad field of research fom "slum tourism", "violence" and "domestic violence" to "gender" and "housing".

\begin{figure}[h!]
\centering
\includegraphics[width=0.7\textwidth]{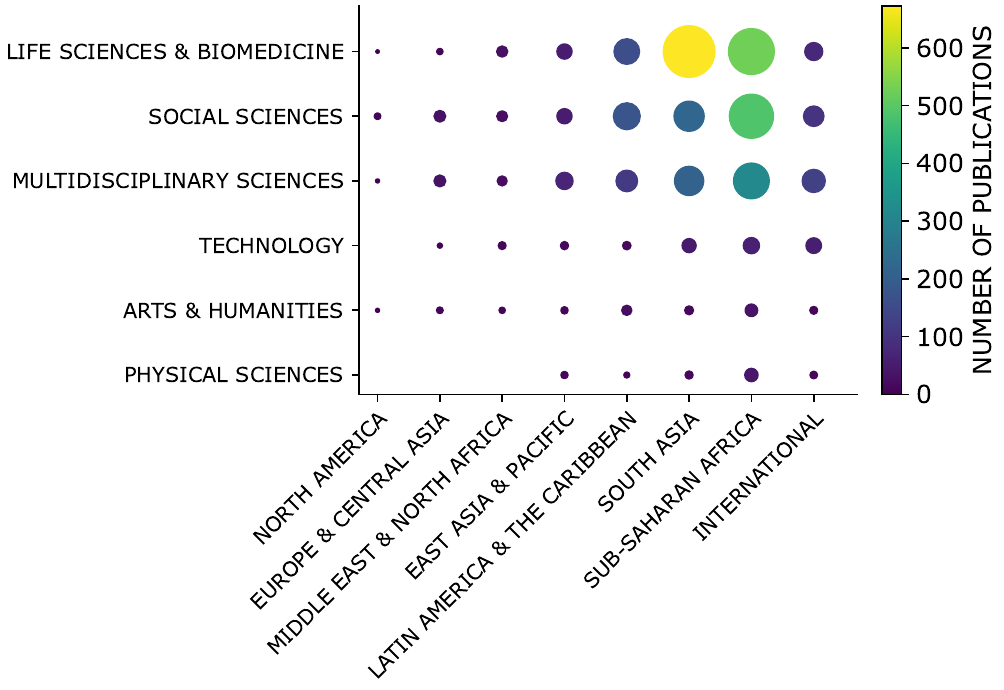}
\caption{The size and color of the scatter plot represent the number of publications in a given research area that contain a researched country in the specified world region.}
\label{fig:ScatterWoSResearchAreaWorldRegion}
\end{figure}

Figure~\ref{fig:ScatterWoSResearchAreaWorldRegion} shows the number of publications for different research areas and research locations, summarized into world regions. Both world regions and research areas are arranged such that the total number of publications within that category increases upward and to the right, respectively. The geographical region \textit{international} is excluded from this order. 

Most of the publications focus on the world regions \textit{Sub-Saharan Africa}, \textit{South Asia} and \textit{Latin America \& Caribbean}, confirming the result of Figure~\ref{fig:reasearched_bar}. In these global regions and nearly all the other regions, the focus of the publications is on the research areas \textit{Life Science \& Biomedicine}, \textit{Social Sciences} and \textit{Multidisciplinary Sciences}. In total, $1528$ publications deal with \textit{Life Science \& Biomedicine}, followed by $1107$ publications in \textit{Social Sciences}.

Looking at the publications classified as \textit{international}, this highest share deals with \textit{Multidisciplinary Sciences}. 

In the geographical region \textit{Latin America \& Caribbean} the research area \textit{Social Sciences} with $177$ publications dominate, followed by $160$ studies on \textit{Life Sciences \& Biomedicine}. This deviates from the overall trend of the dominant research area \textit{Life Sciences \& Biomedicine}. 
For \textit{East Asia \& Pacific} the two research areas \textit{Life Sciences \& Biomedicine} and \textit{Social Sciences} are represented by $54$ publications, while the largest share of publications is concerned with \textit{Multidisciplinary Sciences}. 
Although in total the biggest share of studies deal with countries in \textit{Sub-Saharan Africa}, more studies deal with \textit{Life Science \& Biomedicine} in South Asia than in \textit{Sub-Saharan Africa}. 
For the researched regions \textit{Europe \& Central Asia} and \textit{North America}, most publications were on \textit{Social Science} and not as in \textit{Sub-Saharan Africa} and \textit{South Asia} on \textit{Life Sciences \& Biomedicine}.

The list of the ten most influential journals, shown in Table~\ref{tab:JournalsWithMostCitations_Absolute}, confirms this observation. The journals were ranked using the h-index, as calculated on the basis of the number of citations given in the WoS database. While there are several journals having different aspects of urbanization and the related aspect of deprivation in focus (\textit{Habitat international, Environment and Urbanization, International Journal of Urban and Regional Research, Urban Studies}), most of the other journals deal with issues of health explicitly (\textit{BMC Public Health; Social Science and Medicine; Journal of Urban Health; Journal of Health, Population and Nutrition; Transactions of the Royal Society of Tropical Medicine and Hygiene; American Journal of Tropical Medicine and Hygiene} or \textit{PLOS Neglected tropical diseases}) and implicitly (\textit{PLOS ONE}).

This observation in turn highlights the limitations of our approach, since the amount of citations strongly
depends on the scientific disciplines, as numerous publications have shown (e.g. \cite{Bradshaw.2021}). 

\begin{table}[h!]
    \caption{Top journals based on h-index. We counted all publications for each journal and summed the citations reported in the data set. We also divided the total number of citations by the number of publications, resulting in the values in the forth column.}
    \centering
    \begin{tabular}{l l l l l}
    
        \textbf{Journal} & \textbf{h-index} & \parbox{2.0 cm}{\textbf{Times Cited, All Databases}} & \parbox{2.0 cm}{\textbf{Number of Publications}} & \parbox{2 cm}{\textbf{Citations per publication}}  \\[0.2cm] \hline
        Habitat International & 21 & 1191 & 85 & 14.0\\ 
        BMC Public Health & 20 & 1069 & 69 & 15.5 \\ 
        Environment and Urbanization & 19 & 1076 & 74 & 14.5\\ 
        Social Science and Medicine & 17 & 919 & 33 & 27.8 \\
        PLOS ONE & 16 & 737 & 72 & 10.2 \\  
        Journal of Urban Health & 15 & 555 & 40 & 13.9 \\ 
        Journal of Health, Population and Nutrition & 13 & 493 & 22 & 22.4 \\ 
        \parbox{6cm}{Transactions of the Royal Society of Tropical Medicine and Hygiene} & 12 & 303 & 16 & 18.9 \\ [0.1cm]
        American Journal of Tropical Medicine and Hygiene & 11 & 412 & 23 & 17.9 \\
        International Journal of Urban and Regional Research & 11 & 826 & 19 & 43.5  \\
        PLOS Neglected Tropical Diseases & 11 & 665 & 18 & 36.9  \\ 
        Urban Studies & 11 & 451 & 31 & 14.5 

    \end{tabular}
    \label{tab:JournalsWithMostCitations_Absolute}
\end{table}

\begin{figure}[h!]
\centering
\includegraphics[width=0.98\textwidth]{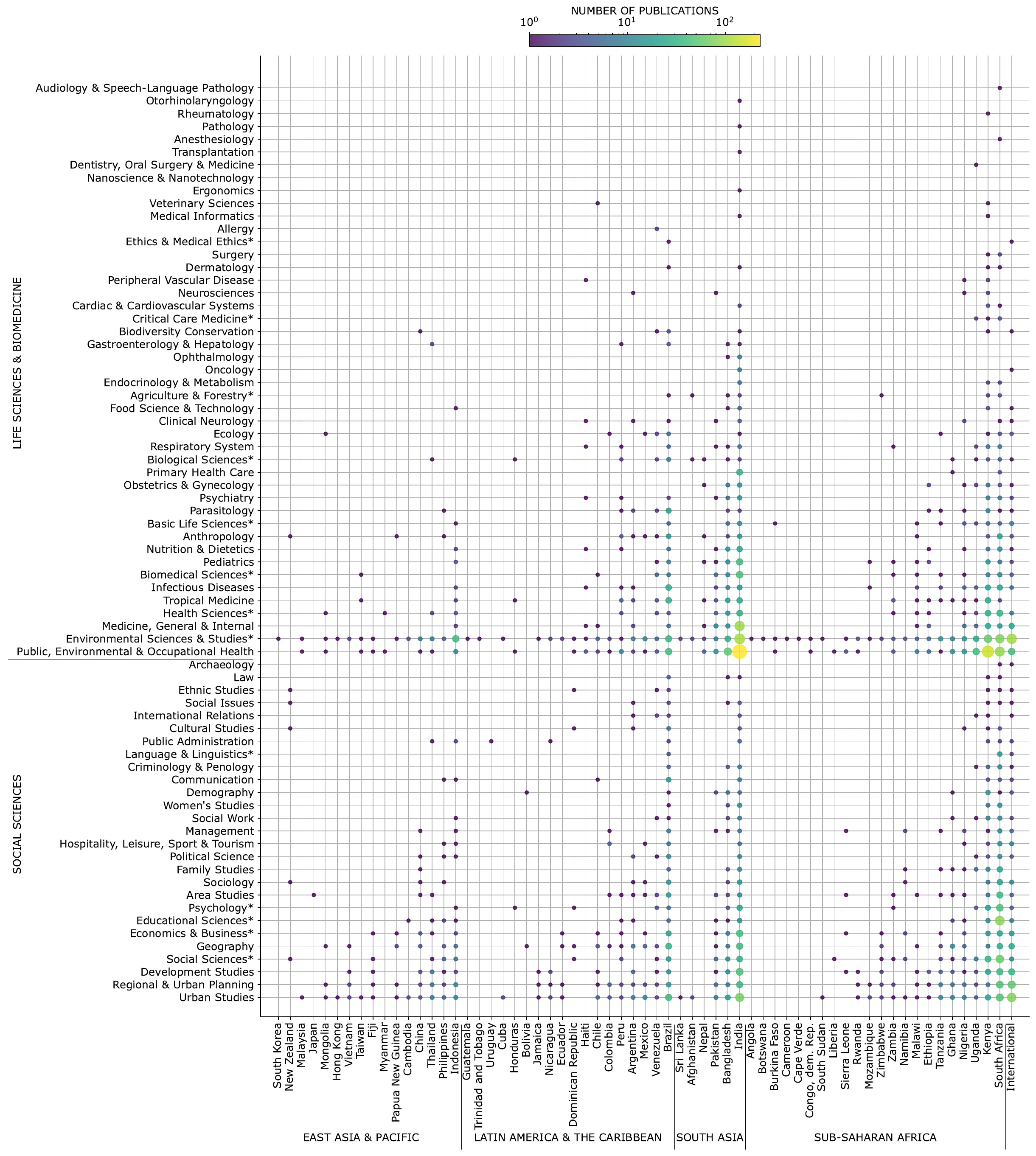}
\caption{Zoom of Figure~\ref{fig:ScatterWoSResearchAreaWorldRegion}: The two research areas \textit{Social Science} and \textit{Life Science \& Biomedicine} and the four world regions \textit{East Asia \& Pacific}, \textit{Latin America \& the Caribbean}, \textit{South Asia} and \textit{Sub-Saharan} \textit{Africa}  are investigated in detail. Furthermore studies classified as \textit{international} are also included. Categories marked with * represent multiple WoS categories (See Sec.~\ref{sec:Material+Methods}).}
\label{fig:ZoomWoSCategoriesResearchedCountries}
\end{figure}

Figure~\ref{fig:ZoomWoSCategoriesResearchedCountries} shows the major geographic locations and research areas of Figure~\ref{fig:ScatterWoSResearchAreaWorldRegion} in more detail. Here all research countries in \textit{East Asia \& Pacific}, \textit{Latin America \& Caribbean},\textit{South Asia} and \textit{Sub-Saharan Africa} are shown. Note that publications assigned to the \textit{Multidisciplinary Science} research area in Figure~\ref{fig:ScatterWoSResearchAreaWorldRegion} are also included here if any of the publication's \textit{WoS categories} match the categories of research area \textit{Life Sciences \& Biomedicine} and \textit{Social Sciences} shown. Furthermore some categories are combined for simplicity \footnote{Such as \textit{Agriculture \& Forestry*} which includes  \textit{Agricultural Economics \& Policy}, \textit{Agriculture, Multidisciplinary} and \textit{Forestry}. \url{https://www.cwts.nl/pdf/nowt_classification_sc.pdf} served as a starting point for summing up the categories}.

It can be seen that the majority of studies often concentrate on specific countries. These specific countries are Brazil for Latin America, India and Bangladesh for South Asia and South Africa and Kenya for Sub-Saharan Africa:

\begin{itemize}
    \item 66\,\% of publications on Latin America \& the Caribbean study Brazil
    \item 76\,\% of publications on South Asia study India
    \item 18\,\% of publications on South Asia study Bangladesh
    \item 47\,\% of publications on Sub-Saharan Africa study South Africa
    \item 30\,\% of publications on Sub-Saharan Africa study Kenya
\end{itemize}

58\,\% of publications on India are connected to the research area \textit{Life Sciences \& Biomedicine}, showing that slums in India are mainly studied from a medical perspective. The same applies to India's neighbor Bangladesh, 54\,\% of research on this country is also associated with this research area. In Kenya, the dominant field of research is also \textit{Life Sciences \& Biomedicine}. In South Africa and Brazil, the majority of publications are in the \textit{Social Sciences} (46\,\% and 40\,\% respectively). South Africa is thus the only country mentioned here that deviates from the dominant research area of its word region. Research on South Africa is therefore more focused on the social sciences than the average research on sub-Saharan Africa. This result is underlined by the results of the keyword analysis presented above.

Looking at the \textit{WoS categories} in Figure~\ref{fig:ZoomWoSCategoriesResearchedCountries}, the focus on public, environmental and occupational health as well as general and internal medicine, tropical medicine and infectious diseases is logical given the huge impact the environment has on human health. It becomes obvious that for certain topics the research focuses on one certain countries. This is the case for \textit{Parasitology}, where 44\,\% of the 45 publications in that category are concerned with Brazil and 20\,\% with Kenya. In contrast, 60\,\% of the publications (139 in total) associated with the title \textit{Educational Sciences*} have South Africa as their country of research. 40\,\% of publications associated with \textit{Psychology*} are researched in South Africa. 42\,\% of publications related to \textit{Pediatrics*} deal with India. It should be mentioned here that these figures could be an effect of the popularity of certain journals in certain countries.

Although the categories in Figure~\ref{fig:ZoomWoSCategoriesResearchedCountries} are based on the publication's journal rather than the publication itself, it seems unrealistic that the relationships described in the previous two paragraphs are a side effect of the popularity of certain journals in certain (geographic) research communities. One aspect that reinforces this is that particularly popular categories such as \textit{Public, Environmental \& Occupational Health} are strongly represented in most heavily researched countries (such as India, South Africa, Kenya, Brazil) and generally no clustering of  \textit{WoS categories} and researching countries is apparent.

It can also be seen, that the countries, in which the highest amount of studies was conducted (e.g. India, Brazil, Kenya or South Africa), show a broad range of perspectives on slums, since nearly all aspects of life sciences, as well as social sciences are represented by at least one study.

Another major result of the analysis of Figure~\ref{fig:ZoomWoSCategoriesResearchedCountries} is that there are a lot of \textit{blind spots} in both, countries and research topics. Countries like Trinidad and Tobago, Sri Lanka, Afghanistan, Angola, Botswana, Burkina Faso or Cameroon all have just a few publications. In addition, there are countries such as Côte d'Ivoire, Madagascar, Mali or El Salvador for which no publication is present in the data set (see Figure~\ref{fig:ScatterSlumPopVsNationalStudies}). This is similar, when the thematic aspects investigated are looked at in more detail. While questions of public, environmental and occupational health are looked at relatively often, there are huge blind spots when it comes to oncology or cardiac and cardiovascular diseases. This is critical, since cancer and cardiovascular diseases are the leading causes of death, also in the Global South and their share is expected to increase further \cite{GBD2019DiseasesandInjuriesCollaborators.2020}.  

\begin{figure}[h!]
\begin{subfigure}[b]{0.5\textwidth}
\centering
\includegraphics[width=1\textwidth]{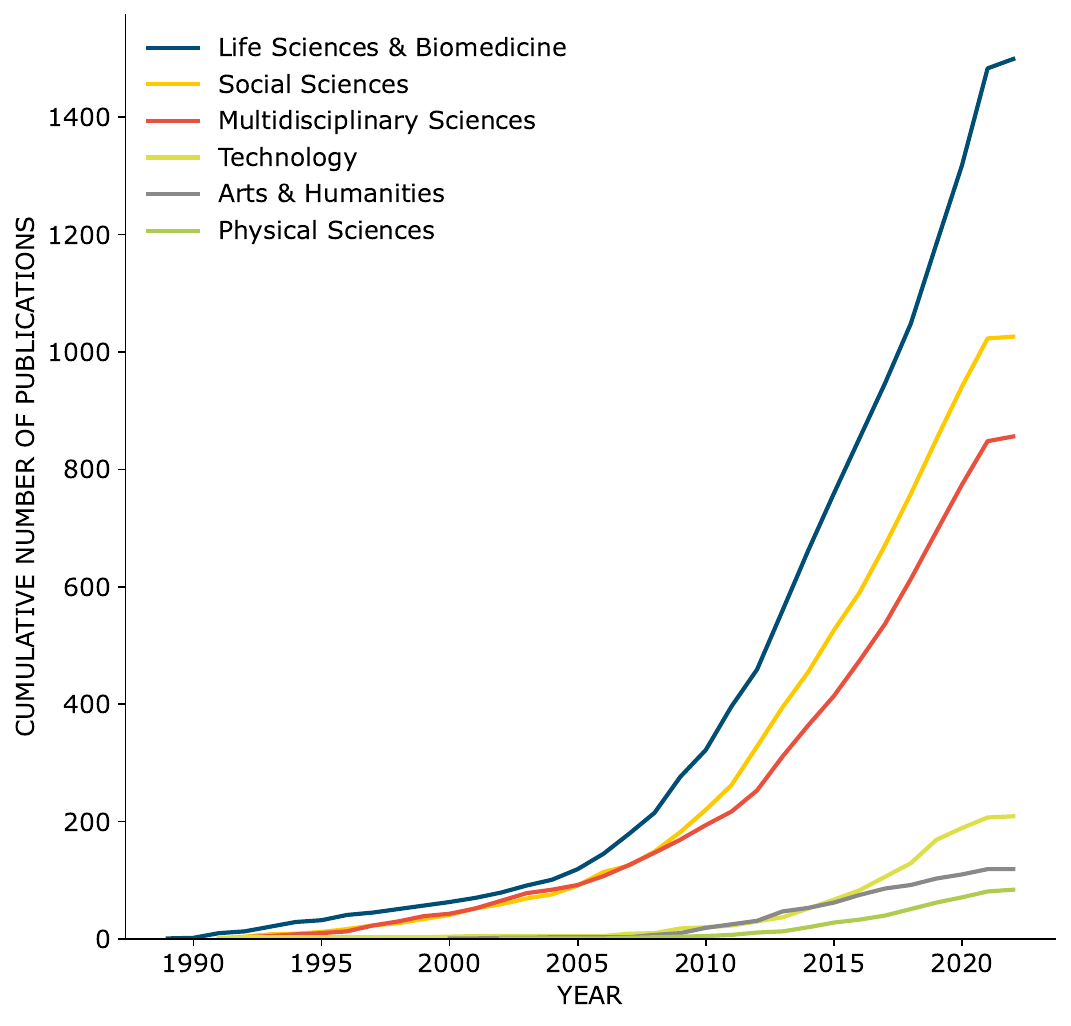}
\caption{Cumulative publications per research area}
\label{fig:temporalResearchArea}
\end{subfigure}
\begin{subfigure}[b]{0.5\textwidth}
\centering
\includegraphics[width=1\textwidth]{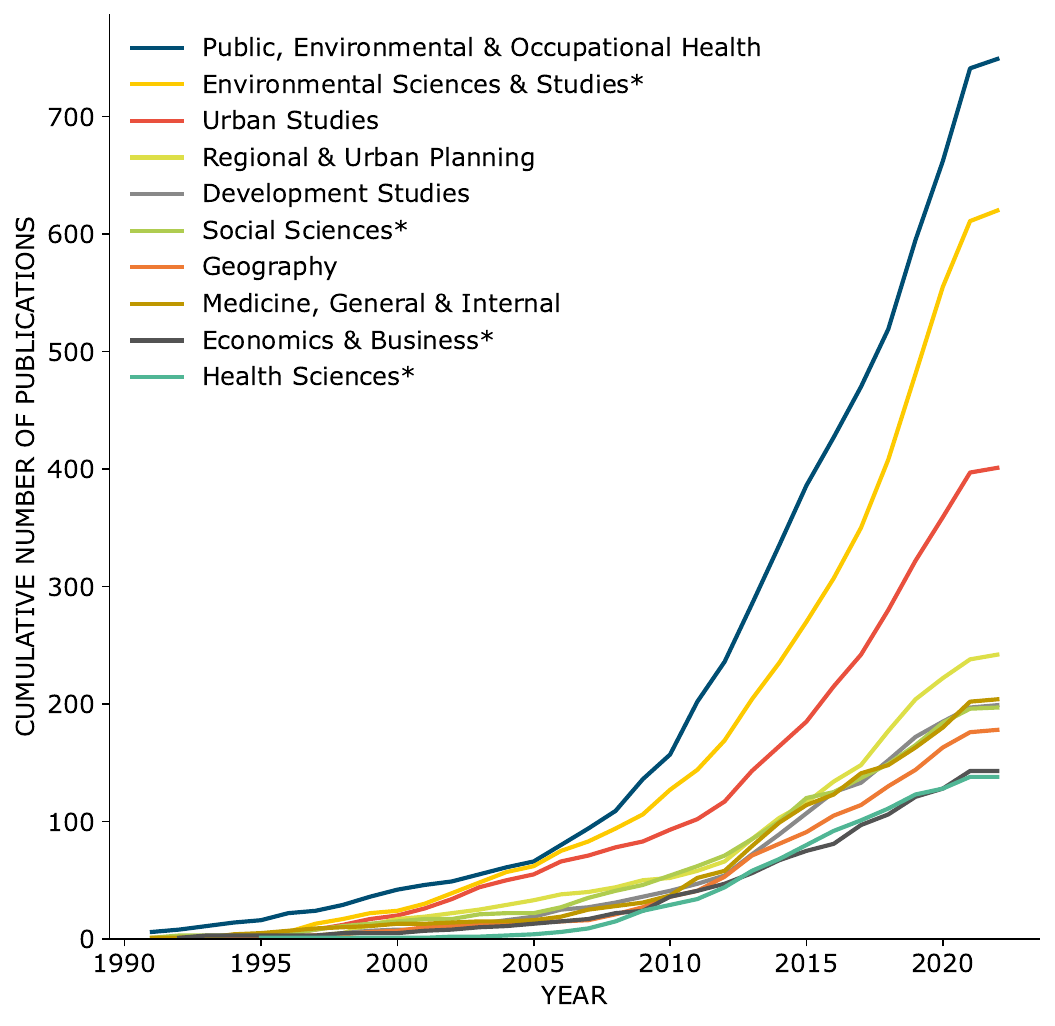}
\caption{Cumulative publications per top 10 WoS Categories}
\label{fig:temporalResearchCategory}
\end{subfigure}
\caption{Temporal development of publications from 1989 to 2022.}
\end{figure}

The temporal development of publications regarding different thematic aspects between 1989 and 2022 is shown in Figure~\ref{fig:temporalResearchArea} and~\ref{fig:temporalResearchCategory}. Since the early 1990s, studies on \textit{Life Sciences \& Biomedicine} dominate the research on slums followed by \textit{Social Sciences}. The third category is \textit{Multidisciplinary Sciences}. In comparison to the categories mentioned before, technological studies are of minor importance. Additionally, studies on environmental sciences are becoming more important in the last years.

\subsection{Research countries \& Countries of research institutions}

Having previously considered the relationship between geographic and thematic research, we now consider the researched countries in relation to the researching countries. We already identified the major researched countries in Figure~\ref{fig:reasearched_bar}. The ten main researching countries based on the institutions given in the address section of a publications metadata are: USA, India, South Africa, United Kingdom, Kenya, Brazil, Netherlands, Bangladesh, Australia and Germany. Note that a publication can have multiple researching countries associated to it.

\begin{figure}[h!]
\centering
\includegraphics[width=0.75\textwidth]{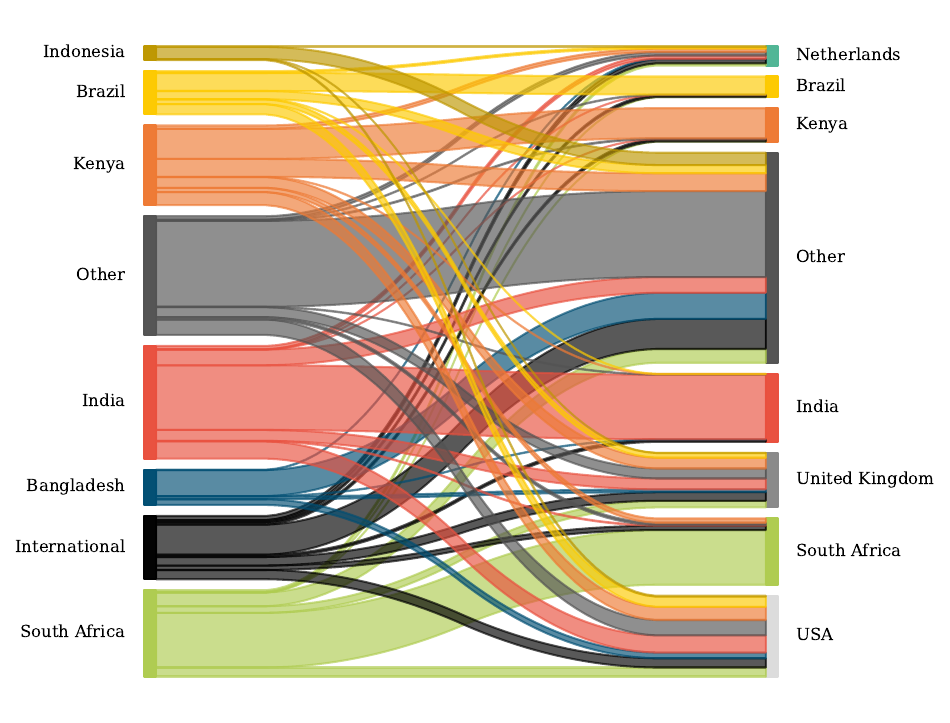}
\caption{The relationship between the six researching countries (right) with the highest number of publications and the six most researched countries (left) is shown here, comparable with the data in Figure~\ref{fig:ResearchedVSInstitutions_Zoom_Other}. All countries not explicitly listed are summarized in "Other".}
\label{fig:ResearchedVSInstitutions_Sankey_Other}
\end{figure}

The combinations of researching countries and researched countries found in our data set are shown in Figure~\ref{fig:ResearchedVSInstitutions_Sankey_Other} where we limited ourselves to the most researched and the most researching countries. A representation of all combinations of researching and researched countries can be found in the Appendix in Figure~\ref{fig:ResearchedVSInstitutions}. Note that in our analysis, the researching country cannot be \textit{international}, as we consider all countries individually that are specified in the authors' affiliations.

The United Kingdom (UK) has the fourth largest number of publications with research institutions located there. The main research institutions based in the UK conduct research in India, Kenya, South Africa, Brazil and Bangladesh, i.e. the intensively researched countries already identified above. A similar pattern is found for the country that most often hosts the research institutions of the publications, the USA. 

There are two major categories of research countries in the data set. The first category represents countries like the USA, the UK, the Netherlands or Australia, with international focus in their research on slums. The second category are nations with a large amount of local research, represented by India, South Africa, Kenya, Brazil or Bangladesh.

Interestingly, the main researching countries with multinational focus conduct little research in Indonesia, a country with a high number of slums. Instead Australia and Japan are the foreign research institutions with the highest presence there. 

Australian research institutions conduct more research on countries in close geographical proximity. Countries like India, Bangladesh, and Indonesia receive greater coverage than Brazil, Kenya, and South Africa. Canada and Sweden are more inclined to publish on Kenya, which differs from the major research powers of the USA and UK. South Africa has a significant presence as a research nation in Kenya and internationally, contrasting with India and Brazil, which conduct little research beyond their own borders.

\section{Discussion}
The results gained by conducting a systematic bibliometric analyses reveal common patterns of research on slums. The research on slums showed a high increase in the last years, with a higher growth rate than in science in general. Bornmann and Mutz \cite{Bornmann.2015} showed for scientific publications independent of the thematic field between 1980 and 2012 a growth rate of 2.9\,\% per year (cf. Table~\ref{tab:Growthrates}) while the growth rate between 1989 and 2021 for publications regarding slums at 10.9\,\% is nearly three times as high.

As expected we can show that most of the studies focus on specific countries, like Kenya, South Africa, India or Brazil. Some of these are also in the list of the countries with the highest ratio of scientific publications to population living in slums (c.f. Figure~\ref{fig:ScatterSlumPopVsNationalStudies}).
We also showed that the majority of publications within the respective world regions are conducted in specific countries, like the ones mentioned above. Nevertheless, there are a lot of countries with high slum population and a low number of or no publications (e.g. Figure~\ref{fig:ScatterSlumPopVsNationalStudies}).

Work that addresses issues of health dominates research on slums and informal settlements. Countless studies discuss the influence that living in slums has on the health of the residents. This relationship has been documented in numerous publications in recent years \cite{Corburn.2012, Lilford.2019, Ezeh.2017}. 
In our database, by far the most studies (770 publications) are associated with the category \textit{Public, Environmental \& Occupational Medicine}. The second and third most popular categories for medical topics are \textit{Medicine, General \& Internal} and \textit{Health Sciences*}.
\cite{Ezeh.2017} pointed out that the health risk of slum dwellers is particularly high for children, it is interesting to see that 91 publications in the dataset are associated with the WoS category \textit{Pediatrics}. This seems low considering the long-term impact of childhood diseases and the amount of studies published in the most common medical categories. Nevertheless, it should be mentioned that keywords related to children frequently appear in the keywords of publications on \textit{Life Sciences \& Biomedicine}.

Two important results from \cite{Ezeh.2017} can also be found in our dataset: Mental health of slum dwellers seems to be little addressed in the literature, as only 111 publications are linked to the category \textit{psychology*} in our database. 
Cancer research is represented by 9 publications in the category \textit{oncology}, which is consistent with Ezeh's finding that non-communicable diseases rarely appear in publications on the health of slum dwellers.

Social constellations and the view of slum dwellers on living together, as well as the view from the Global North on the phenomenon and the related constellation of power are discussed in multiple publications (e.g. \cite{Roy.2011}).

For studies in the category \textit{Physical Sciences} both, most cited publications and author keywords show a similar pattern. The most investigated issues are water related topics like flooding or sanitation and questions of resilience. Although climate change is one of the biggest, if not the biggest challenge of the beginning 21st century, the frequency of studies using this topic as keyword is comparably low.

When looking at the category \textit{Technology} we find that classification of slums using remote sensing or geoinformation systems is a topic occuring often in highly cited journals. This is also reflected in the wordcloud of the author keywords. It is interesting, that although fire spread is also an often mentioned keyword, but up to now, now influential publication discussed this topic. We also didn't found any holistic review of the state of the art, although fire spreads are a major health risk for slum dwellers \cite{Ezeh.2017}. ICT4D (Information and Communications Technologies for Development), an other often mentioned key word reflects on a movement to make the access to digital technologies more accessible in the Global South.

These findings on technological and infrastructural questions draw a more nuanced picture than previous studies, reflecting on slums and modern technological concepts \cite{Friesen.2022}.

It is also important to mention, that a quantitative study on a global scale has several limitations, which have to be kept in mind, when the results are looked at. 

\subsection{Limitations}
\label{sec:limitation}
Since we conduct a combination of a scoping and a bibliometrical study, we rely on the number of publication as metric to assess the amount of research conducted in a specific country. Using this metric leads to a bias, since publication strategies vary a lot between different disciplines. Since the most cited journals and those with the highest impact factors come from the medical domain, it is clear that these papers and therefore the connected issues dominate our statistics. Thereby, as usual in scoping studies \cite{arksey2005scoping}, no differentiation was made between comprehensive review articles or short conference papers. Nevertheless, the statistics give a trend of publication strategies in different countries.
A similar limitation emerges with the number of slum population as a weighting factor. It would also be possible to use the share of urban population living in slums as metric, but this in turn would lower the importance of big countries like India or Nigeria and the heterogeneity of their slum populations, when the share would be equal to the ones of much smaller countries, like e.g. the Maldives.

This leads to another bias resulting from aggregating the publications on a country level. Although our statistics in Figure~\ref{fig:studperinh} suggest that the density of studies per resident is high for India, it may be that these studies are limited to only a certain part of the country. For example, Friesen et al. \cite{Friesen.2020b} show on the example of medical studies in slums that the bulk of the work in India relates only to Mumbai and the bulk of the work in Kenya relates only to Nairobi, even though many slums are present in the rest of those countries as well. Thinking of the very high number of slum dwellers especially in India, further investigations are necessary to analyse the spatial distribution and thereby the coverage of the knowledge in more detail.

Another major limitation of our study is the list of terms used in the literature search. Our search was limited to the same list as in the paper by Ezeh et al. \cite{Ezeh.2017}. The use of the mentioned terminology shifts the focus of our study towards the Global South, despite the fact that urban deprivation can also be found regions of the Global North, as Kraff et al. \cite{Kraff.2022} impressively showed. Therefore, the focus of the paper is much more on deprived settlements in the Global South. Although we include the term deprived areas in our literature search, we removed manually some of the related publications in the United Kingdom and Sweden, since it don't describe a different category of settlement type. If a block of houses has water and sanitation services, residents have a lease, and the number of residents is limited, then the five dimensions the UN requires for categorization as a slum are all not met. At the same time, countless studies show severe deprivation in these areas. This is similar for the above-mentioned term \textit{barrio}. It is used to describe settlements in Latin America fulfilling the conditions to be classified as slum according to the UN. Socially deprived areas in cities in the USA are also referred to as \textit{barrio} in publications we found during our WoS search. Although the distinction with settlements that meet the UN definition of slums may be possible, we excluded these publications with research reference to the USA from our analysis. We are aware that this practice itself introduces bias into our analysis, as it does not adequately represent research on deprived areas in the USA.
In addition to the term \textit{barrio}, the terms \textit{township} (used for research on South Africa) and \textit{favela} (used for research on Brazil) can also be used in a different sense than the term "slum".

The problem can be captured by two quotes of Ludwig Wittgenstein \cite{Wittgenstein.2022}: (i) "The limit of our language is the limit of our world." and (ii) "The meaning of a word is its use in the language." 
It is a well known fact, that due to the ambiguities in slum definitions, it is difficult to compare different concepts and one has to be careful to not compare apples with oranges \cite{Lilford.2019}. A more integrated framework, such as the one proposed by Abascal et al. \cite{Abascal.2022} is still needed to be implemented.

By excluding non-English publications from our database, we may be omitting a significant number of scientific publications on slums, which also affects the results of our analysis. One effect of this restriction might be the low number of publications in Western Sub-Saharan Africa. For instance, there could be a number of relevant publications in French but nonetheless, we deliberately excluded non-English publications as it was not possible for us to make a profound evaluation of non-English abstracts.
One indication of this weakness of our database is the very low number of English-language publications with France as a research country in the data set.
Another limitation is the fact, that we limited our search query on the titles of publications. Thereby, we probably missed important publications on the topic, without any of our search terms, like the well known work on urban informality of Roy~\cite{roy.2005}. Additionally, it has to be noticed, that it could be that our method excludes certain scientific disciplines that deal with the topic of slums but publish in book chapters and thus fall outside our search pattern.

All of these limitations are complex and should be kept in mind when the quantitative findings of our analysis are considered. It has to be kept in mind, that the goal of bibliometrical analyses and scoping reviews is to map the field and show open research gaps and do not qualitativly assess the literature. Nevertheless, the analyses performed above can just be seen as a starting point for more holistic and interdisciplinary approaches to uncover the problems and inequalities, and their consequences in more detail. 

\subsection{Directions for research}
There are several questions for further research that this study could be the starting point for. Based on our analysis we highlight four different exemplary directions for further research on slums:
        
\subsubsection*{Analysis and increase effort to decrease blind spots (spatial and thematic)}
Firstly, our results highlight the necessity for research to overcome the many thematic and spatial blind spots in research on slums. Our knowledge on slums should not only rely on studies conducted in specific countries representing different world regions. Our results highlight the low density of studies per population in Western sub-Saharan Africa. It is also important to search for the reasons for these blind spots. One possible question would be to what extend accessibility and societal conditions shape the scientific research. Do recent conflicts, such as in Sudan or long lasting civil wars correlate with scientific output in the respective region?

It is also important to look more deeply at a regional level to see similarities and differences in living conditions of slums within the same country or the same city. These specific analyses can in turn highlight the specific necessities the dwellers have and can lead to adequate and specific actions, reducing deprivation.

Looking at thematic issues, it has to be emphasized that medical research on slums should not only focus on public and environmental health for example, but also investigate the risk of cancer or cardiovascular diseases. Both of these classes of diseases are main causes of death world wide, but their specifics are not well understood when it comes to slums \cite{Friesen.2020b}. 

Although research on technological questions are diverse and cover a high range of topics, we found that there are still open topics, like the need for a comprehensive review on fire spread in informal settlements. Another major point is the need for studies (especially comprehensive systematic reviews) on climate change and slums, since these area are often prone to natural disasters which are expected to occur more often in the following years.

\subsubsection*{Analyse connection between countries and thematic research}
Another crucial area for future research involves examining how the location of a study relates to its chosen theme. We need to explore whether there are observable connections between specific types of countries and the thematic areas they tend to focus on. If such connections exist, it is essential to investigate the underlying reasons. For example, it is evident that research on diseases like malaria and tropical illnesses is particularly concentrated in Sub-Saharan Africa (SSA). This concentration can be attributed, at least in part, to the substantial burden these diseases place on the local population, making it logical to prioritize research in this geographic region.

It could also be analysed which topics are most relevant in the respective countries. One first step could be to create lists like Table~\ref{tab:TopPupblications} for different countries.

\subsubsection*{Analyse connections between researching and researched countries}
Another open question deals with the relationship between researching and researched countries. Does the researching country have an influence on the thematic focus of the research in the researched country? To what extent does the historical involvement, e.g. in the context of the colonial period, play a role? To what extend do intergovernmental partnerships influence funding of collaborative research and thereby the scientific output?

\subsubsection*{Shift from dichotomous perspective to a more holistic one}
Slums and informal settlements are dichotomous concepts and often related to constellations of power \cite{Roy.2011,Mayne.2017,GILBERT.2007}. While research was often and is still conducted from a top-down perspective, future research should explicitly involve slum dwellers and their view into scientific research that pertains to them \cite{Roy.2011}. Therefore, participatory frameworks should rather be used than top-down approaches by foreign researchers to gain a more comprehensive picture \cite{Abascal.2022}. Looking at the temporal development, it is clear that this trend is already there and in a majority of the most researched countries the domestic and collaborative studies have a higher growth than the foreign studies, c.f. Figure~\ref{fig:mixed_foreign_native_over_years}. 

Finally, it is worth noting in this context that a more general thematic perspective can also broaden research geographically. Instead of looking for slums and urban poverty primarily in the Global South, holistic frameworks are needed that examine disadvantage in different dimensions \cite{Abascal.2022}. This would lead to a differentiated picture of urban deprivation, in which different perspectives could lead to a holistic picture.

\section{Conclusion}
According to United Nations estimates, about one in four urban dwellers lives in a so-called slum. 
Through bibliometric analysis on a global scale, we were able to track past trends - both thematically and spatially - in research on this category of settlement. In addition to the existing metadata, our analysis included information about where the studies were conducted.
Through our bibliometric analysis at the global level, where we were able to add information on the location of the study in addition to existing metadata, we were able to trace previous trends (both thematically and spatially) in research on this class of settlement.

Since research on slums has so far focused mainly on specific countries such as India, South Arica, Kenya or Brazil, large parts and thus a considerable proportion of the slum population remain unconsidered in the scientific literature. 

Although further in-depth research, especially of a qualitative nature and inform of systematic literature reviews, is needed to examine the relationships observed here in more detail, this study provides an overview of the current state of affairs and can serve as a starting point for establishing further perspectives in research on slums.

\printcredits

\section*{Competing interests}
There is NO competing interest. 

\section*{Data availability}
The Excel file containing the list of publications and the geographic assignment is attached as Supplemental Material.

\appendix
\section{Appendix}

\subsection{Search Methodology}
\label{sec:searchmethodology}
To focus especially on slums, we used the list of names for slums from Ezeh et al. \cite{Ezeh.2017}. We conducted the search on March 4th, 2022 using the following search terms:
((TI = (baladi*  OR boras de miseria*  OR barraca*  OR barrio marginal*  OR barrio*  OR bidonville*  OR brarek*  OR bustee*  OR chalis*  OR chereka bete*  OR dagatan*  OR deprived area* OR favela*  OR galoos*  OR gecekondu*  OR hrushebi*  OR informal settlement*  OR ishash*  OR karyan*  OR katras*  OR looban*  OR loteamento*  OR medina achouaia*  OR morro*  OR mudun safi*  OR musseque*  OR shanty town*  OR slum*  OR solares*  OR tanake*  OR taudis*  OR township*  OR tugurio*  OR udukku*  OR umjondolo*  OR watta*  OR zopadpattis*)  NOT  TI=(SLUMP*  OR  munro township*  OR slumgu*))) AND LANGUAGE: (English) Refined by: [excluding] PUBLICATION TITLES: ( AAPG BULLETIN AMERICAN ASSOCIATION OF PETROLEUM GEOLOGISTS ) 

This search gained 10,073 results. We removed the ones without abstract resulting in 5,543 publications. Two of the authors screened all publications, checked if they were related to slums.
Since there were a lot of studies analysing deprivation in the UK or in Sweden, we excluded the studies, if none of the other terms were used. In addition, the term \textit{barrio} in studies related to research in the USA mostly refers to Latino neighborhoods, which is why studies with the research country USA and the term \textit{barrio} are not part of the analysed data set.

\FloatBarrier

\subsection{Further Results}

\begin{table}[h!]
    \caption{Top institutions with an h-index greater than or equal to 15. The information on citations is also included in the Web of Science data base we downloaded and thereby restricted to all citations before the date the data set was downloaded (March 4th, 2022).}
    \centering
    \begin{tabular}{l l l l l l}
    
        \textbf{Institution} & \textbf{Country} & \textbf{h-index} & \parbox{2 cm}{\textbf{Times Cited, All Databases}} & \parbox{2 cm}{\textbf{Number of Publications}} & \parbox{1.7 cm}{\textbf{Citations per publication}} \\[0.2cm] \hline
        \parbox{5cm}{African Population and Health Research Center} & Kenya & 28 &2558 & 155 & 16.5 \\[0.3cm]
        University of Cape Town & South Africa & 23& 1598 & 129 & 12.4 \\[0.1cm]
        \parbox{5cm}{The London School of Hygiene \& Tropical Medicine} & UK & 22 & 1319 & 84 & 15.7 \\[0.3cm] 
        Johns Hopkins University & USA & 21 & 1629 & 67 & 24.3 \\[0.1cm] 
        University of the Witwatersrand & South Africa & 19 & 1379 & 113 & 12.2 \\[0.1cm] 
        Stellenbosch University & South Africa & 17 & 986 & 96 & 10.3 \\[0.1cm] 
        All India Institutes of Medical Sciences & India & 15 & 1012 & 41 & 24.7 \\[0.1cm] 
        University of the Western Cape &  South Africa  & 15 & 741 & 46 & 16.1\\[0.1cm]
        Harvard University & USA	& 15 & 709 &	39	& 18.2 \\[0.1cm] 
        Makerere University & Uganda & 15 & 738 & 63 & 11.7\\[0.1cm]
    \end{tabular}
    \label{tab:InstitutionsWithMostCitations_Absolute}
\end{table}

\begin{table}[h!]
    \caption{Relative frequency of the five most frequently used keywords without geographical reference, shown in Figure~\ref{fig:KeywordsLifeScience}. 1273  publications with keywords are considered.}
    \centering
    \begin{tabular}{ll}
         \textbf{keyword}&\textbf{frequency} \\\hline
        children      &        3.61\,\%\\
        prevalence     &       2.99\,\%\\
        hiv             &      2.83\,\%\\
        sanitation       &     2.75\,\%\\
        malnutrition      &    2.60\,\%\\
    \end{tabular}
    \label{tab:frequencyKeywordsLifeScience}
\end{table}

\begin{figure}[h!]
\centering
\includegraphics[width=0.8\textwidth]{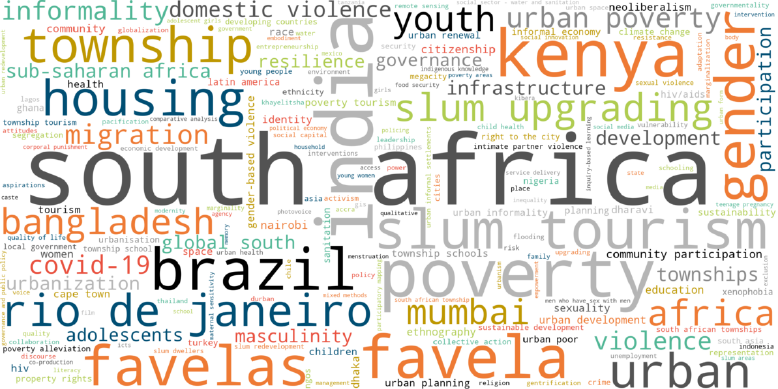}
\caption{Word cloud consisting of author keywords for publications with research area \textit{Social Sciences}. Generic and obvious keywords such as "urban slum" and "informal settlement" are not taken into account.}
\label{fig:KeywordsSocialScience}
\end{figure}

\begin{table}[h!]
    \caption{Relative frequency of the six most frequently used keywords without geographical reference, shown in fig.~\ref{fig:KeywordsSocialScience}. 878 publications with keywords are considered.}
    \centering
    \begin{tabular}{ll}
         \textbf{keyword}&\textbf{frequency} \\\hline
            poverty       &  4.55\,\%\\
            favela         & 3.53\,\%\\
            favelas        & 3.08\,\%\\
            slum tourism   & 3.08\,\%\\
            gender         & 3.08\,\%\\
            township       & 2.39\,\%\\
            housing        & 2.39\,\%\\
    \end{tabular}
    \label{tab:frequencyKeywordsSocialScience}
\end{table}

\begin{figure}[h!]
\centering
\includegraphics[width=0.8\textwidth]{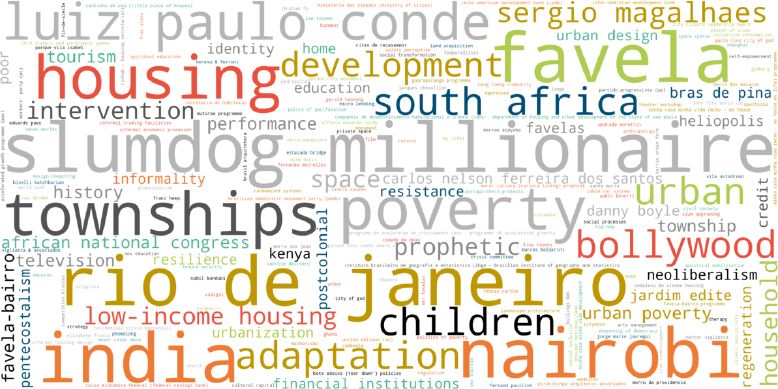}
\caption{Word cloud consisting of author keywords for publications with research area \textit{Arts \& Humanities}. Generic and obvious keywords such as "urban slum" and "informal settlement" are not taken into account.}
\label{fig:KeywordsArts}
\end{figure}

\begin{table}[h!]
    \caption{Relative frequency of the twelve most frequently used keywords without geographical reference, shown in fig.~\ref{fig:KeywordsArts}. 94 publications with keywords are considered.}
    \centering
    \begin{tabular}{ll}
         \textbf{keyword}&\textbf{frequency} \\\hline
            slumdog millionaire &   4.26\,\%\\
            favela               &  4.26\,\%\\
            housing              &  3.19\,\%\\
            poverty               & 3.19\,\%\\
            townships             & 3.19\,\%\\
            luiz paulo conde      & 3.19\,\%\\
            adaptation            & 3.19\,\%\\
            bollywood             & 3.19\,\%\\
            children              & 3.19\,\%\\
            urban                 & 3.19\,\%\\
            development           & 3.19\,\%\\
    \end{tabular}
    \label{tab:frequencyKeywordsArts}
\end{table}

\begin{figure}[h!]
\centering
\includegraphics[width=0.8\textwidth]{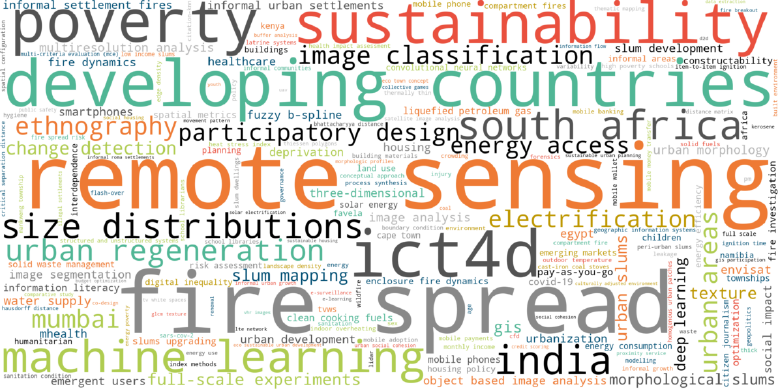}
\caption{Word cloud consisting of author keywords for publications with research area \textit{Technology}. Generic and obvious keywords such as "urban slum" and "informal settlement" are not taken into account.}
\label{fig:KeywordsTech}
\end{figure}

\begin{table}[h!]
    \caption{Relative frequency of the seven most frequently used keywords without geographical reference, shown in fig.~\ref{fig:KeywordsTech}. 172 publications with keywords are considered.}
    \centering
    \begin{tabular}{ll}
         \textbf{keyword}&\textbf{frequency} \\\hline
            remote sensing      &    5.23\,\% \\
            fire spread          &   4.65\,\% \\
            developing countries  &  3.49\,\% \\
            ict4d                 &  3.49\,\% \\
            poverty               &  2.91\,\% \\
            sustainability         & 2.91\,\% \\
            machine learning       & 2.91\,\% \\
    \end{tabular}
    \label{tab:frequencyKeywordsTech}
\end{table}

\begin{figure}[h!]
\centering
\includegraphics[width=0.8\textwidth]{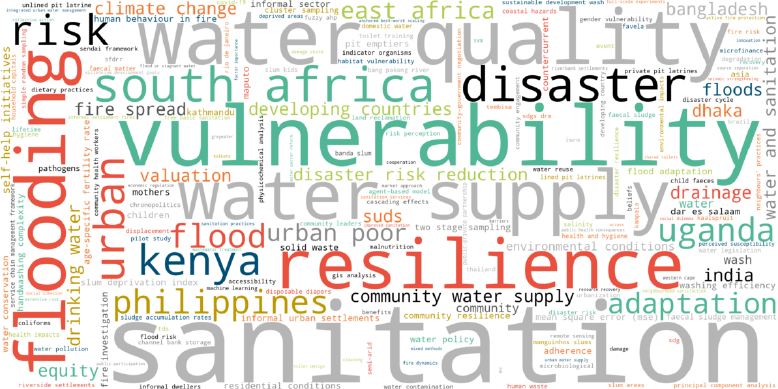}
\caption{Word cloud consisting of author keywords for publications with research area \textit{Physical Sciences}. Generic and obvious keywords such as "urban slum" and "informal settlement" are not taken into account.}
\label{fig:KeywordsPhysi}
\end{figure}

\begin{table}[h!]
    \caption{Relative frequency of the eight most frequently used keywords without geographical reference, shown in fig.~\ref{fig:KeywordsPhysi}. 78 publications with keywords are considered.}
    \centering
    \begin{tabular}{ll}
         \textbf{keyword}&\textbf{frequency} \\\hline
            sanitation     &  15.38\,\%\\
            vulnerability   &  7.69\,\%\\
            water quality   &  6.41\,\%\\
            resilience      &  5.13\,\%\\
            water supply    &  5.13\,\%\\
            flooding        &  5.13\,\%\\
            disaster        &  5.13\,\%\\
    \end{tabular}
    \label{tab:frequencyKeywordsPhysi}
\end{table}

\begin{figure}[h!]
\centering
\includegraphics[width=0.8\textwidth]{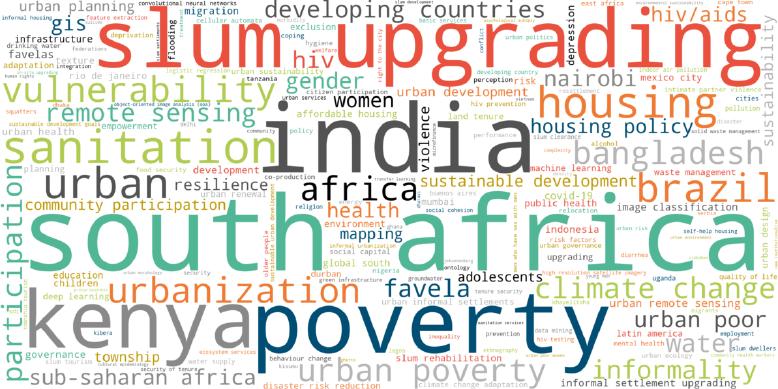}
\caption{Word cloud consisting of author keywords for publications with research area \textit{Multidisciplinary Sciences}. Generic and obvious keywords such as "urban slum" and "informal settlement" are not taken into account.}
\label{fig:KeywordsMulti}
\end{figure}

\begin{table}[h!]
    \caption{Relative frequency of the eight most frequently used keywords without geographical reference, shown in fig.~\ref{fig:KeywordsMulti}. 689 publications with keywords are considered.}
    \centering
    \begin{tabular}{ll}
         \textbf{keyword}&\textbf{frequency} \\ \hline
        poverty          & 4.35\,\%\\
        slum upgrading   & 3.77\,\%\\
        sanitation       & 3.48\,\%\\
        housing          & 3.05\,\%\\
        urban            & 2.90\,\%\\

    \end{tabular}
    \label{tab:frequencyKeywordsMulti}
\end{table}

\begin{figure}[h!]
\centering
\includegraphics[width=0.65\textwidth]{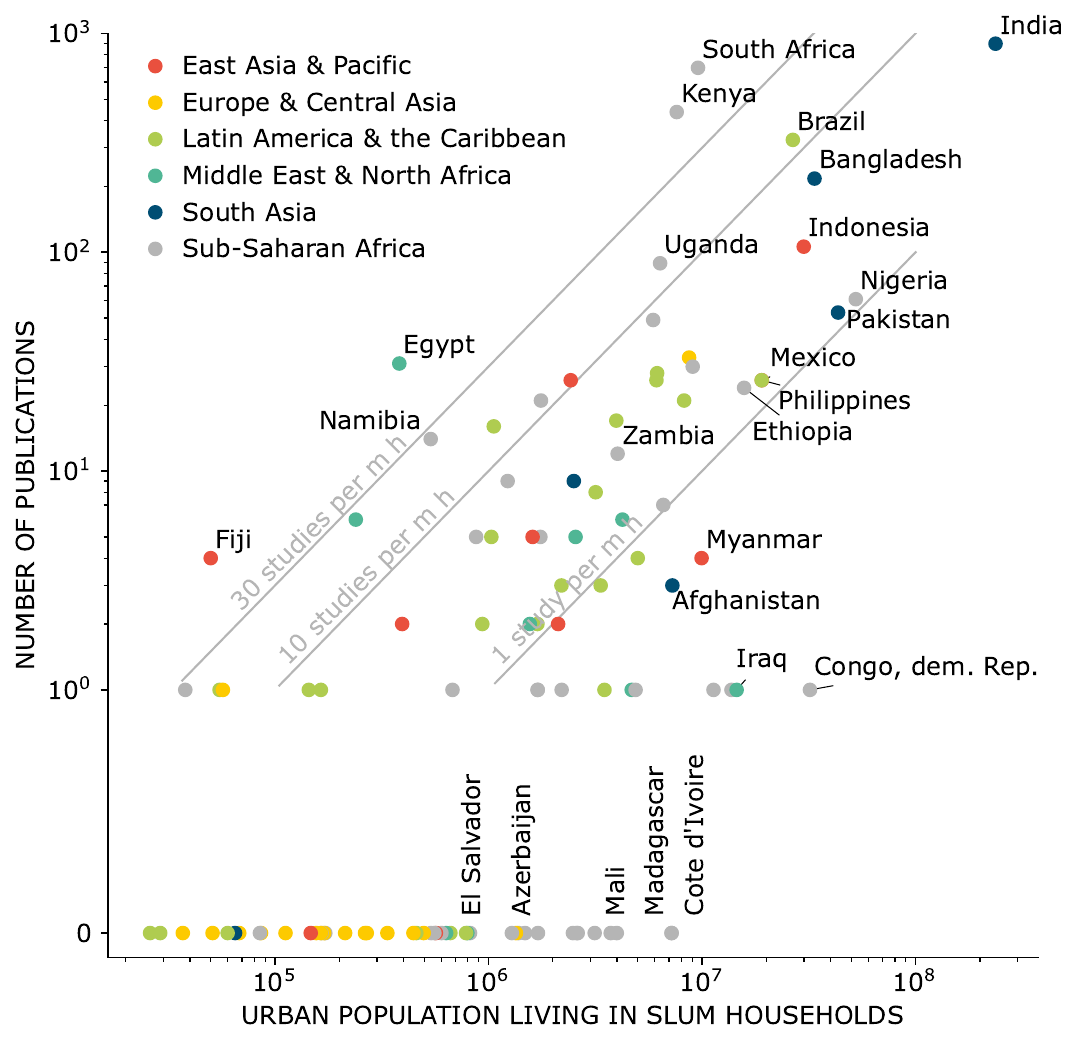}
\caption{Shows same data as Figure~\ref{fig:studperinh}}
\label{fig:ScatterSlumPopVsNationalStudies}
\end{figure}

\begin{figure}[h!]
\centering
\includegraphics[width=0.65\textwidth]{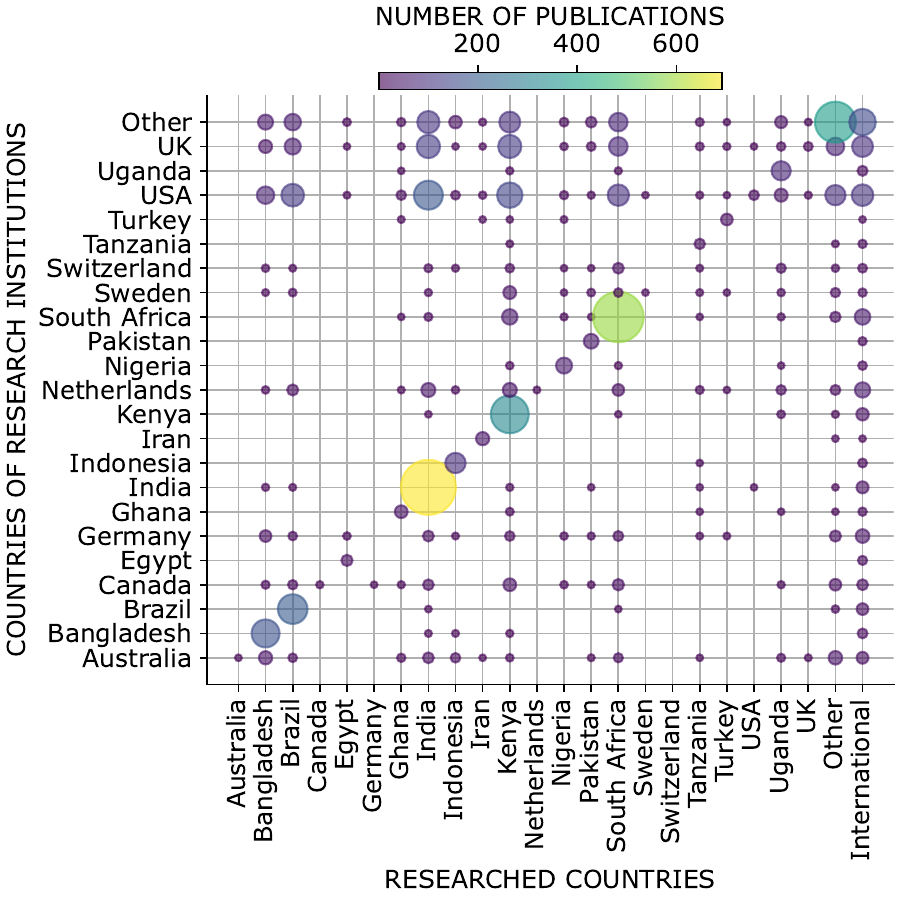}
\caption{The 15 researching countries with the highest amount of publications and the 15 most researched countries. Since some of the countries are in both categories (one of the countries with the most research and one of the most researched countries, e.g. India), there are 23 countries. All other data are summarized in 'Other'. 
The complete Figure can be seen in the Appendix (Figure~\ref{fig:ResearchedVSInstitutions}).}
\label{fig:ResearchedVSInstitutions_Zoom_Other}
\end{figure}

\begin{figure}[h!]
\centering
\includegraphics[width=\textwidth]{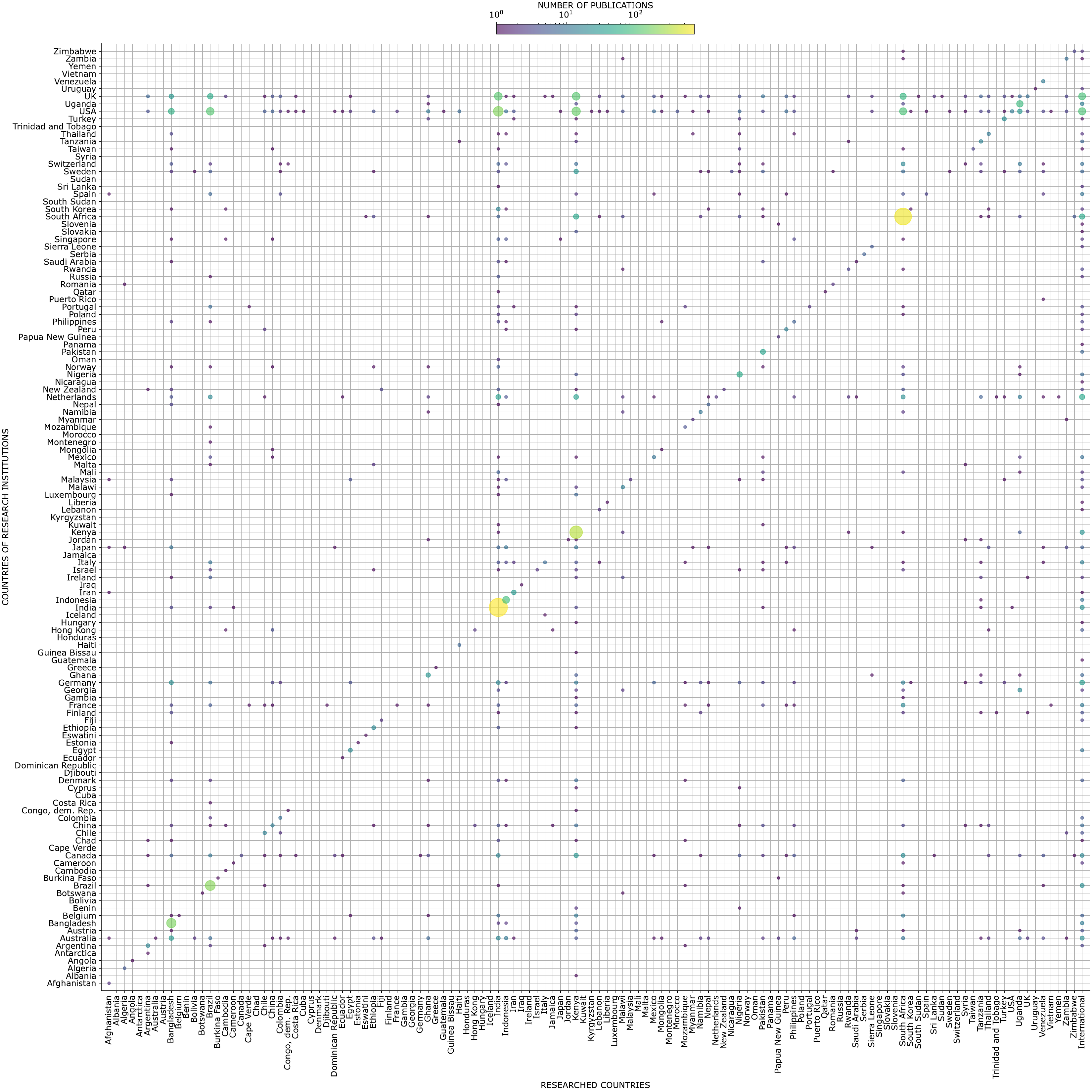}
\caption{The x-axis shows the researched countries, the y-axis shows the location of the researching institutions. Note the logarithmic scaling in the legend.}
\label{fig:ResearchedVSInstitutions}
\end{figure}

\begin{sidewaystable}
\small
\def\arraystretch{2.5}
    \caption{The ten most cited publications in the research area \textit{Life Sciences \& Biomedicine} in the data set. The column "Times Cited" corresponds to the column "Times Cited, All Databases" in WoS metadata, "Year" corresponds to "Publication Year". WoS categories are separated by ';' and researching countries are sorted alphabetically.}
    \centering
    \resizebox{\textwidth}{!}{
    \begin{tabular}{l l l l l l l}
        
       \parbox{0.8cm}{\textbf{Times Cited}} & \textbf{Article Title} & \textbf{Authors} & \textbf{Year} & \textbf{WoS Categories} & \textbf{Researched Country} & \parbox{2.7cm}{\textbf{Researching Countries}} \\  [0.2cm]  \hline

        451 & \parbox{6cm}{HIV testing attitudes, AIDS stigma, and voluntary HIV counselling and testing in a black township in Cape Town, South Africa} \cite{Kalichman.2003} & \parbox{5cm}{Kalichman, Sc; Simbayi, Lc}& 2003 & Infectious Diseases & South Africa & South Africa, USA
        \\ [0.5cm] \hline
        
        303 & \parbox{6cm}{Exclusive breastfeeding reduces acute respiratory infection and diarrhea deaths among infants in Dhaka slums} \cite{Arifeen.2001} & \parbox{5cm}{Arifeen, S; Black, Re; Antelman, G; Baqui, A; Caulfield, L; Becker, S} & 2001 & Pediatrics & Bangladesh & USA
        \\ [0.5cm] \hline
        
        248 & \parbox{6cm}{High prevalence of diabetes, obesity and dyslipidaemia in urban slum population in northern India} \cite{Misra.2001} & \parbox{5cm}{Misra, A; Pandey, Rm; Devi, Jr; Sharma, R; Vikram, Nk; Khanna, N} & 2001 & \parbox{4cm}{Endocrinology \& Metabolism; Nutrition \& Dietetics} & India & India
        \\ [0.5cm] \hline
        
        234 & \parbox{6cm}{Effect of anthelmintic treatment on the allergic reactivity of children in a tropical slum} \cite{Lynch.1993} & \parbox{5cm}{Lynch, Nr; Hagel, I; Perez, M; Diprisco, Mc; Lopez, R; Alvarez, N} & 1993 & Allergy; Immunology & Venezuela & Venezuela
        \\ [0.5cm] \hline
        
        216 &\parbox{6cm}{The state of emergency obstetric care services in Nairobi informal settlements and environs: Results from a maternity health facility survey} \cite{Ziraba.2009} & \parbox{5cm}{Ziraba, Ak; Mills, S; Madise, N; Saliku, T; Fotso, Jc} & 2009 & \parbox{4cm}{Health Care Sciences \& Services} & Kenya & Kenya, UK, USA, 
        \\ [0.5cm] \hline
        
        207 & \parbox{6cm}{Impact of Environment and Social Gradient on Leptospira Infection in Urban Slums} \cite{Reis.2008} & \parbox{5cm}{Reis, Rb; Ribeiro, Gs; Felzemburgh, Rdm; Santana, Fs; Mohr, S; Melendez, Axto; Queiroz, A; Santos, Ac; Ravines, Rr; Tassinari, Ws; Carvalho, Ms; Reis, Mg; Ko, Ai} & 2008 & \parbox{4cm}{Infectious Diseases; Parasitology; Tropical Medicine} & Brazil & Brazil, USA 
        \\ [0.5cm] \hline
        
        141 & \parbox{6cm}{Bacteremic typhoid fever in children in an urban slum, Bangladesh} \cite{Brooks.2005} & \parbox{5cm}{Brooks, Wa; Hossain, A; Goswami, D; Sharmeen, At; Nahar, K; Alam, K; Ahmed, N; Naheed, A; Nair, Gb; Luby, S; Breiman, Rf} & 2005 & \parbox{4cm}{Immunology; Infectious Diseases} & Bangladesh & Bangladesh
        \\ [0.5cm] \hline
        
        129 & \parbox{6cm}{Determinants of health care seeking for childhood illnesses in Nairobi slums} \cite{Taffa.2005} & \parbox{5cm}{Taffa, N; Chepngeno, G} & 2005 & \parbox{4cm}{Public, Environmental \& Occupational Health; Tropical Medicine} & Kenya & Kenya
        \\ [0.5cm] \hline
        
        124 & \parbox{6cm}{Effect of mother's education on child's nutritional status in the slums of Nairobi} \cite{Abuya.2012} & \parbox{5cm}{Abuya, Ba; Ciera, J; Kimani-Murage, E} & 2012 & Pediatrics & Kenya & Kenya
        \\ [0.5cm] \hline
        
        118 & \parbox{6cm}{Epidemiologic and clinical characteristics of acute diarrhea with emphasis on Entamoeba histolytica infections in preschool children in an urban slum of Dhaka, Bangladesh} \cite{SACK.2003} & \parbox{5cm}{Haque, R; Mondal, D; Kirkpatrick, Bd; Akther, S; Farr, Bm; Sack, Rb; Petri, Wa} & 2003 & \parbox{4cm}{Public, Environmental \& Occupational Health; Tropical Medicine} & Bangladesh & Bangladesh, USA
        
    \end{tabular}
    }
    \label{tab:TopPupblicationsLifeScience}
\end{sidewaystable}

\begin{sidewaystable}
\small
\def\arraystretch{2.5}
    \caption{The ten most cited publications in the research area \textit{Social Sciences} in the data set. The column "Times Cited" corresponds to the column "Times Cited, All Databases" in WoS metadata, "Year" corresponds to "Publication Year". WoS categories are separated by ';' and researching countries are sorted alphabetically.}
    \centering
    \resizebox{\textwidth}{!}{
    \begin{tabular}{l l l l l l l}
        
       \parbox{0.8cm}{\textbf{Times Cited}} & \textbf{Article Title} & \textbf{Authors} & \textbf{Year} & \textbf{WoS Categories} & \textbf{Researched Country} & \parbox{2.7cm}{\textbf{Researching Countries}} \\  [0.2cm]  \hline

         395 & \parbox{6cm}{Slumdog Cities: Rethinking Subaltern Urbanism} \cite{Roy.2011} & Roy, A & 2011 & \parbox{4cm}{Geography; Regional \& Urban Planning; Urban Studies} & International & USA
        \\ [0.5cm] \hline

        166 & \parbox{6cm}{Making the best of a bad situation: Satisfaction in the slums of Calcutta} \cite{Biswas.2021} & \parbox{5cm}{Biswas-Diener, R; Diener, E} & 2001 & \parbox{4cm}{Social Sciences, Interdisciplinary; Sociology} & India & USA 
        \\ [0.5cm] \hline
        
        164 & \parbox{6cm}{The return of the slum: Does language matter?} \cite{GILBERT.2007} & Gilbert, A & 2007 & \parbox{4cm}{Geography; Regional \& Urban Planning; Urban Studies} & International &  UK
        \\ [0.5cm] \hline
        
        126 & \parbox{6cm}{Property rights and investment in urban slums} \cite{Field.2005} & Field, E & 2005 & Economics & International & USA 
        \\ [0.5cm] \hline
        
        115 & \parbox{6cm}{Cooperation, trust, and social capital in Southeast Asian urban slums} \cite{Carpenter.2004} & \parbox{5cm}{Carpenter, JP; Daniere, AG; Takahashi, LM} & 2004 & Economics & International & Canada, USA 
        \\ [0.5cm] \hline
        
        108 & \parbox{6cm}{Sexual risk-taking in the slums of Nairobi, Kenya, 1993-98} \cite{Zulu.2002} & \parbox{5cm}{Zulu, EM; Dodoo, FNA; Chika-Ezeh, A} & 2002 & Demography & Kenya & Kenya, USA 
        \\ [0.5cm] \hline
        
        92 & \parbox{6cm}{'I'm used to it now': experiences of homophobia among queer youth in South African township schools} \cite{Msibi.2012} & Msibi, T & 2012 & \parbox{4cm}{Education \& Educational Research} & South Africa & South Africa 
        \\ [0.5cm] \hline
        
        91 & \parbox{6cm}{Towards a material ethnography of linguistic landscape: Multilingualism, mobility and space in a South African township} \cite{Stroud.2009} & \parbox{5cm}{Stroud, C; Mpendukana, S} & 2009 & Linguistics & South Africa & South Africa
        \\ [0.5cm] \hline
        
        89 & \parbox{6cm}{'Smuggling the vernacular into the classroom': conflicts and tensions in classroom codeswitching in township/rural schools in South Africa} \cite{Probyn.2009} & Probyn, M & 2009 & \parbox{4cm}{Education \& Educational Research; Linguistics; Language \& Linguistics} & South Africa & South Africa 
        \\ [0.5cm] \hline
        
        84 & \parbox{6cm}{Residential Satisfaction In China's Informal Settlements: A Case Study Of Beijing, Shanghai, And Guangzhou} \cite{Li.2013} & Li, ZG; Wu, FL & 2013 & \parbox{4cm}{Geography; Urban Studies} & China & China, UK 
        \\ [0.5cm] \hline

    \end{tabular}
    }
    \label{tab:TopPupblicationsSocialSciences}
\end{sidewaystable}

\begin{sidewaystable}
\small
\def\arraystretch{2.5}
    \caption{The ten most cited publications in the research area \textit{Multidisciplinary Sciences} in the data set. The column "Times Cited" corresponds to the column "Times Cited, All Databases" in WoS metadata, "Year" corresponds to "Publication Year". WoS categories are separated by ';' and researching countries are sorted alphabetically.}
    \centering
    \resizebox{0.9\textwidth}{!}{
    \begin{tabular}{l l l l l l l}
        
       \parbox{0.8cm}{\textbf{Times Cited}} & \textbf{Article Title} & \textbf{Authors} & \textbf{Year} & \textbf{WoS Categories} & \textbf{Researched Country} & \parbox{2.7cm}{\textbf{Researching Countries}} \\  [0.2cm]  \hline 

       339 & \parbox{6cm}{'I think condoms are good but, aai, I hate those things': condom use among adolescents and young people in a Southern African township} \cite{MacPhail.2001} & \parbox{4cm}{MacPhail, C; Campbell, C}  & 2001 & \parbox{4cm}{Public, Environmental \& Occupational Health; Social Sciences, Biomedical} & South Africa &  South Africa, UK
       \\ [0.5cm] \hline
         
        114& \parbox{6cm}{Prevalence and Factors Associated with Intestinal Parasitic Infection among Children in an Urban Slum of Karachi} \cite{Mehraj.2008} & \parbox{4cm}{Mehraj, V; Hatcher, J; Akhtar, S; Rafique, G; Beg, MA}  & 2008  & \parbox{4cm}{Multidisciplinary Sciences} & Pakistan & France, Kuwait, Pakistan
        \\ [0.5cm] \hline

        108 & \parbox{6cm}{Sanitation in Mumbai's informal settlements: state, 'slum', and infrastructure} \cite{McFarlane.2008} & McFarlane, C & 2008 & \parbox{4cm}{Environmental Studies; Geography} & India & UK 
        \\ [0.5cm] \hline
        
        105 & \parbox{6cm}{An ontology of slums for image-based classification} \cite{Kohli.2012} & \parbox{4cm}{Kohli, D; Sliuzas, R; Kerle, N; Stein, A} & 2012 &  \parbox{4cm}{Computer Science, Interdisciplinary Applications; Engineering, Environmental; Environmental Studies; Geography; Operations Research \& Management Science; Regional \& Urban Planning} & International & Netherlands
        \\ [0.5cm] \hline
        
        101 & \parbox{6cm}{Population-Based Incidence of Typhoid Fever in an Urban Informal Settlement and a Rural Area in Kenya: Implications for Typhoid Vaccine Use in Africa} \cite{Breiman.2012} & \parbox{4cm}{Breiman, RF; Cosmas, L; Njuguna, H; Audi, A; Olack, B; Ochieng, JB; Wamola, N; Bigogo, GM; Awiti, G; Tabu, CW; Burke, H; Williamson, J; Oundo, JO; Mintz, ED; Feikin, DR} & 2012 & Multidisciplinary Sciences & Kenya & Kenya, USA
        \\ [0.5cm] \hline
        
        91 & \parbox{6cm}{When HIV-prevention messages and gender norms clash: The impact of domestic violence on women's HIV risk in slums of Chennai, India} \cite{Go.2003} & \parbox{4cm}{Go, VF; Sethulakshmi, CJ; Bentley, ME; Sivaram, S; Srikrishnan, AK; Solomon, S; Celentano, DD} & 2003 &  \parbox{4cm}{Public, Environmental \& Occupational Health; Social Sciences, Biomedical} & India & India, USA 
       \\ [0.5cm] \hline
        
        91 & \parbox{6cm}{Alcohol and sexual risk behavior among men who have sex with men in South African township communities} \cite{Lane.2008} & \parbox{4cm}{Lane, T; Shade, SB; McIntyre, J; Morin, SF} & 2008 &  \parbox{4cm}{Public, Environmental \& Occupational Health; Social Sciences, Biomedical} & South Africa & South Africa, USA 
        \\ [0.5cm] \hline
        
        87 & \parbox{6cm}{Sexual assault history and risks for sexually transmitted infections among women in an African township in Cape Town, South Africa} \cite{Kalichman.2004} & \parbox{4cm}{Kalichman, SC; Simbayi, LC} & 2004 &  \parbox{4cm}{Health Policy \& Services; Public, Environmental \& Occupational Health; Psychology, Multidisciplinary; Respiratory System; Social Sciences, Biomedical} & South Africa & South Africa, USA 
        \\ [0.5cm] \hline
        
        82 & \parbox{6cm}{Effects of Micro-Enterprise Services on HIV Risk Behaviour Among Female Sex Workers in Kenya's Urban Slums} \cite{Odek.2009} & \parbox{4cm}{Odek, WO; Busza, J; Morris, CN; Cleland, J; Ngugi, EN; Ferguson, AG} & 2009 &  \parbox{4cm}{Public, Environmental \& Occupational Health; Social Sciences, Biomedical} & Kenya & Canada, Kenya, UK 
        \\ [0.5cm] \hline

        81 & \parbox{6cm}{Spatio-temporal modelling of informal settlement development in Sancaktepe district, Istanbul, Turkey} \cite{Dubovyk.2011} & \parbox{4cm}{Dubovyk, O; Sliuzas, R; Flacke, J} & 2011 &  \parbox{4cm}{Geography, Physical; Geosciences, Multidisciplinary; Remote Sensing; Imaging Science \& Photographic Technology} & Turkey & Netherlands 
        
    \end{tabular}
    }
    \label{tab:TopPupblicationsMultidisciplinary}
\end{sidewaystable}

\begin{sidewaystable}
\small
\def\arraystretch{2.5}
    \caption{The ten most cited publications in the research area \textit{Technology} in the data set. The column "Times Cited" corresponds to the column "Times Cited, All Databases" in WoS metadata, "Year" corresponds to "Publication Year". WoS categories are separated by ';' and researching countries are sorted alphabetically.}
    \centering
    \resizebox{\textwidth}{!}{
    \begin{tabular}{l l l l l l l}
        
       \parbox{0.8cm}{\textbf{Times Cited}} & \textbf{Article Title} & \textbf{Authors} & \textbf{Year} & \textbf{WoS Categories} & \textbf{Researched Country} & \parbox{2.7cm}{\textbf{Researching Countries}} \\  [0.2cm]  \hline 

        83 & \parbox{6cm}{Slums from Space - 15 Years of Slum Mapping Using Remote Sensing} \cite{Kuffer.2016} & \parbox{4cm}{Kuffer, M; Pfeffer, K; Sliuzas, R} & 2016 & \parbox{4cm}{Remote Sensing} & International & Netherlands 
        \\ [0.5cm] \hline
        
        58 & \parbox{6cm}{Mobility, Poverty, and Gender: Travel 'Choices' of Slum Residents in Nairobi, Kenya} \cite{Salon.2010} & \parbox{4cm}{Salon, D; Gulyani, S} & 2010 & Transportation & Kenya & USA 
        \\ [0.5cm] \hline
        
        47 & \parbox{6cm}{Transferability of Object-Oriented Image Analysis Methods for Slum Identification} \cite{Kohli.2013} & \parbox{4cm}{Kohli, D; Warwadekar, P; Kerle, N; Sliuzas, R; Stein, A} & 2013 & Remote Sensing & India & Netherlands 
        \\ [0.5cm] \hline
        
        28 & \parbox{6cm}{A semi-automated approach for extracting buildings from QuickBird imagery applied to informal settlement mapping} \cite{Mayunga.2007} & \parbox{4cm}{Mayunga, SD; Coleman, DJ; Zhang, Y} & 2007 & \parbox{4cm}{Remote Sensing; Imaging Science\& Photographic Technology} & International & Canada
         \\ [0.5cm] \hline
         
        27 & \parbox{6cm}{Detection of Informal Settlements from VHR Images Using Convolutional Neural Networks} \cite{Mboga.2017} & \parbox{4cm}{Mboga, N; Persello, C; Bergado, JR; Stein, A} & 2017 & Remote Sensing & Tanzania & Netherlands 
        \\ [0.5cm] \hline
        
        26 & \parbox{6cm}{Use of IKONOS satellite data to identify informal settlements in Dehradun, India} \cite{Jain.2007} & Jain, S & 2007 & \parbox{4cm}{Remote Sensing; Imaging Science \& Photographic Technology} & India & India
        \\ [0.5cm] \hline
        
        20 & \parbox{6cm}{The Mobile Phone Store Ecology in a Mumbai Slum Community: Hybrid Networks for Enterprise} \cite{Rangaswamy.2010} & \parbox{4cm}{Rangaswamy, N; Nair, S} & 2010 & Information Science \& Library Science & India & India, USA
        \\ [0.5cm] \hline
        
        19 & \parbox{6cm}{Online Favela: The Use of Social Media by the Marginalized in Brazil} \cite{Nemer.2016} & Nemer, D & 2016 & \parbox{4cm}{Information Science \& Library Science} & Brazil & India, USA
        \\ [0.5cm] \hline
        
        19 & \parbox{6cm}{Using Multi-criteria Evaluation and GIS for Flood Risk Analysis in Informal Settlements of Cape Town: The Case of Graveyard Pond} \cite{Musungu.2012} &\parbox{4cm}{Musungu, K; Motala, S; Smit, J} & 2012 & Remote Sensing & South Africa & South Africa
        \\ [0.5cm] \hline
        
        19 & \parbox{6cm}{Detecting social groups from space - Assessment of remote sensing-based mapped morphological slums using income data} \cite{Wurm.2018} & \parbox{4cm}{Wurm, M; Taubenbock, H} & 2018 & \parbox{4cm}{Remote Sensing; Imaging Science \& Photographic Technology} & Brazil & Germany

    \end{tabular}
    }
    \label{tab:TopPupblicationsTechnology}
\end{sidewaystable}

\begin{sidewaystable}
\small
\def\arraystretch{2.5}
    \caption{The twelve most cited publications in the research area \textit{Arts \& Humanities} in the data set. The column "Times Cited" corresponds to the column "Times Cited, All Databases" in WoS metadata, "Year" corresponds to "Publication Year". WoS categories are separated by ';' and researching countries are sorted alphabetically.}
    \centering
    \resizebox{0.8\textwidth}{!}{
    \begin{tabular}{l l l l l l l}
        \parbox{0.8cm}{\textbf{Times Cited}} & \textbf{Article Title} & \textbf{Authors} & \textbf{Year} & \textbf{WoS Categories} & \textbf{Researched Country} & \parbox{2.7cm}{\textbf{Researching Countries}} \\  [0.2cm]  \hline 
        
        31 & \parbox{6cm}{Building the city of the future: Visions and experiences of modernity in Ghana's akosombo township} \cite{MIESCHER.2012} & Miescher, SF & 2012 & History & Ghana & USA 
        \\ [0.5cm] \hline
        
        16 & \parbox{6cm}{From 'slum clearance' to 'revitalisation': planning, expertise and moral regulation in Toronto's Regent Park} \cite{James.2010} & James, RK & 2010 &  \parbox{4cm}{Architecture; History; History Of Social Sciences} & Canada & Canada
        \\ [0.5cm] \hline
        
        14 & \parbox{6cm}{‘Bloodhounds as Detectives’ Dogs, Slum Stench and Late-Victorian Murder Investigation} \cite{Pemberton.2013} & Pemberton, N & 2013 & History & UK & UK
        \\ [0.5cm] \hline
        
        14 & \parbox{6cm}{Showcasing India Unshining: Film Tourism in Danny Boyle's Slumdog Millionaire} \cite{Mendes.2010} & Mendes, AC & 2010 & Art & India & Portugal
         \\ [0.5cm] \hline
        
        14 & \parbox{6cm}{Slum Clearance, Privatization and Residualization: the Practices and Politics of Council Housing in Mid-twentieth-century England} \cite{Jones.2010} & Jones, B & 2010 & History & UK & UK
        \\ [0.5cm] \hline
        
        11 & \parbox{6cm}{Unmarried Muslim youth and sex education in the bustees of Kolkata} \cite{Chakraborty.2010} & Chakraborty, K & 2010 & Asian Studies & India & Australia
        \\ [0.5cm] \hline
        
        10 & \parbox{6cm}{Al Asmakh historic district in Doha, Qatar: from an urban slum to living heritage} \cite{Boussaa.2014} & Boussaa, D & 2014 & Architecture & Qatar & Qatar
        \\ [0.5cm] \hline
        
        8 & \parbox{6cm}{Slums, Squats, or Hutments? Constructing and Deconstructing an In-Between Space in Modern Shanghai (1926-65)} \cite{ChristianHenriot.2012} & Henriot, C & 2012 & History & China & France
        \\ [0.5cm] \hline
        
        8 & \parbox{6cm}{Langa Township in the 1920s - an (extra)ordinary Garden Suburb} \cite{NicholasCoetzer.2009} & Coetzer, N & 2009 & Art & South Africa & South Africa 
        \\ [0.5cm] \hline
        
        8 & \parbox{6cm}{The lending practices of township micro-lenders and their impact on the low-income households in South Africa: A case study for Mamelodi township} \cite{Mashigo.2012} & Mashigo, P & 2012 & History & South Africa & South Africa
        \\ [0.5cm] \hline

        8 & \parbox{6cm}{THE BOND OF EDUCATION: GENDER, THE VALUE OF CHILDREN, AND THE MAKING OF UMLAZI TOWNSHIP IN 1960s SOUTH AFRICA} \cite{Hunter.2014} & Hunter, M & 2014 & History & South Africa & Canada, South Africa
        \\ [0.5cm] \hline

        8 & \parbox{6cm}{'Umkhonto we Sizwe, We are Waiting for You': The ANC and the Township Uprising, September 1984-September 1985} \cite{Simpson.2009} & Simpson, T & 2009 & History & South Africa & South Africa
    \end{tabular}
    }
    \label{tab:TopPupblicationsArtsHumanities}
\end{sidewaystable}

\begin{sidewaystable}
\small
\def\arraystretch{2.5}
    \caption{The ten most cited publications in the research area \textit{Physical Sciences} in the data set. The column "Times Cited" corresponds to the column "Times Cited, All Databases" in WoS metadata, "Year" corresponds to "Publication Year". WoS categories are separated by ';' and researching countries are sorted alphabetically.}
    \centering
    \resizebox{\textwidth}{!}{
    \begin{tabular}{l l l l l l l}
        \parbox{0.8cm}{\textbf{Times Cited}} & \textbf{Article Title} & \textbf{Authors} & \textbf{Year} & \textbf{WoS Categories} & \textbf{Researched Country} & \parbox{2.7cm}{\textbf{Researching Countries}} \\  [0.2cm]  \hline 
        
        59 & \parbox{6cm}{Floods in megacity environments: vulnerability and coping strategies of slum dwellers in Dhaka/Bangladesh} \cite{Braun.2011} & Braun, B; Assheuer, T & 2011 & \parbox{4cm}{Geosciences, Multidisciplinary; Meteorology \& Atmospheric Sciences; Water Resources} & Bangladesh & Germany 
        \\ [0.5cm] \hline
        
        58 & \parbox{6cm}{Communal sanitation alternatives for slums: A case study of Kibera, Kenya} \cite{Schouten.2010} & \parbox{4cm}{Schouten, MAC; Mathenge, RW} & 2010 & \parbox{4cm}{Geosciences, Multidisciplinary; Meteorology \& Atmospheric Sciences; Water Resources} & Kenya & Kenya, Netherlands 
        \\ [0.5cm] \hline
        
        47 & \parbox{6cm}{Flood risk assessment for informal settlements} \cite{Risi.2013} & \parbox{4cm}{De Risi, R; Jalayer, F; De Paola, F; Iervolino, I; Giugni, M; Topa, ME; Mbuya, E; Kyessi, A; Manfredi, G; Gasparini, P}  & 2013 & \parbox{4cm}{Geosciences, Multidisciplinary; Meteorology \& Atmospheric Sciences; Water Resources} & Tanzania & Italy, Tanzania
        \\ [0.5cm] \hline
        
        38 & \parbox{6cm}{Effect of sanitation facilities, domestic solid waste disposal and hygiene practices on water quality in Malawi's urban poor areas: a case study of South Lunzu Township in the city of Blantyre} \cite{Palamuleni.2002} & Palamuleni, LG & 2002 & \parbox{4cm}{Geosciences, Multidisciplinary; Meteorology \& Atmospheric Sciences; Water Resources} & Malawi & Malawi
        \\ [0.5cm] \hline
         
        22 & \parbox{6cm}{Thirsty slums in African cities: household water insecurity in urban informal settlements of Lilongwe, Malawi} \cite{Adams.2018} & Adams, EA & 2018 & Water Resources & Malawi & Georgia, USA
        \\ [0.5cm] \hline
        
        22 & \parbox{6cm}{Can the vulnerable be resilient? Co-existence of vulnerability and disaster resilience: Informal settlements in the Philippines} \cite{Usamah.2014} & \parbox{4cm}{Usamah, M; Handmer, J; Mitchell, D; Ahmed, I} & 2014 & \parbox{4cm}{Geosciences, Multidisciplinary; Meteorology \& Atmospheric Sciences; Water Resources} & Philippines & Australia
        \\ [0.5cm] \hline
        
        19 & \parbox{6cm}{Urban farming in the informal settlements of Atteridgeville, Pretoria, South Africa} \cite{vanAverbeke.2009} & van Averbeke, W & 2007 & Water Resources & South Africa & South Africa 
         \\ [0.5cm] \hline
         
        14 & \parbox{6cm}{Towards measurable resilience: A novel framework tool for the assessment of resilience levels in slums} \cite{Woolf.2016} & \parbox{4cm}{Woolf, S; Twigg, J; Parikh, P; Karaoglou, A; Cheaib, T} & 2016 & \parbox{4cm}{Geosciences, Multidisciplinary; Meteorology \& Atmospheric Sciences; Water Resources} & Kenya & Canada, Chad, UK
        \\ [0.5cm] \hline
        
        14 & \parbox{6cm}{Disaster risk construction in the progressive consolidation of informal settlements: Iquique and Puerto Montt (Chile) case studies} \cite{Castro.2015} & \parbox{4cm}{Castro, CP; Ibarra, I; Lukas, M; Ortiz, J; Sarmiento, JP} & 2015 & \parbox{4cm}{Geosciences, Multidisciplinary; Meteorology \& Atmospheric Sciences; Water Resources} & Chile & Chile, USA
        \\ [0.5cm] \hline
        
        12 & \parbox{6cm}{Blue Diversion: a new approach to sanitation in informal settlements} \cite{Larsen.2015} & \parbox{4cm}{Larsen, TA; Gebauer, H; Grundl, H; Kunzle, R; Luthi, C; Messmer, U; Morgenroth, E; Niwagaba, CB; Ranner, B} & 2015 & \parbox{4cm}{Water Resources} & International & Austria, Switzerland, Uganda

    \end{tabular}
    }
    \label{tab:TopPupblicationsPhysicalSciences}
\end{sidewaystable}

\begin{table}
    \caption{Parameters for the fitting of the overall temporal development using the function $m \cdot \exp{(b \cdot (x-x_\text{min})}$.}
    \centering
    \begin{tabular}{llll}
         &\multicolumn{2}{l}{\textbf{Research on slums}} & \textbf{Research in general} \cite{Bornmann.2015}\\
        \textbf{Years} & 1989-2021 & 1990-2012 & 1980-2012 \\\hline
        $x_\text{min}$ & 1989 & 1990 & 1980\\ 
        $b$ & 0.109 & 0.18 & 0.029 \\ 
        $m$ & 12.811 & 3.46 & 702880 \\ 
    \end{tabular}
    \label{tab:Growthrates}
\end{table}

\begin{figure}[h!]
\centering
\includegraphics[width=0.5\textwidth]{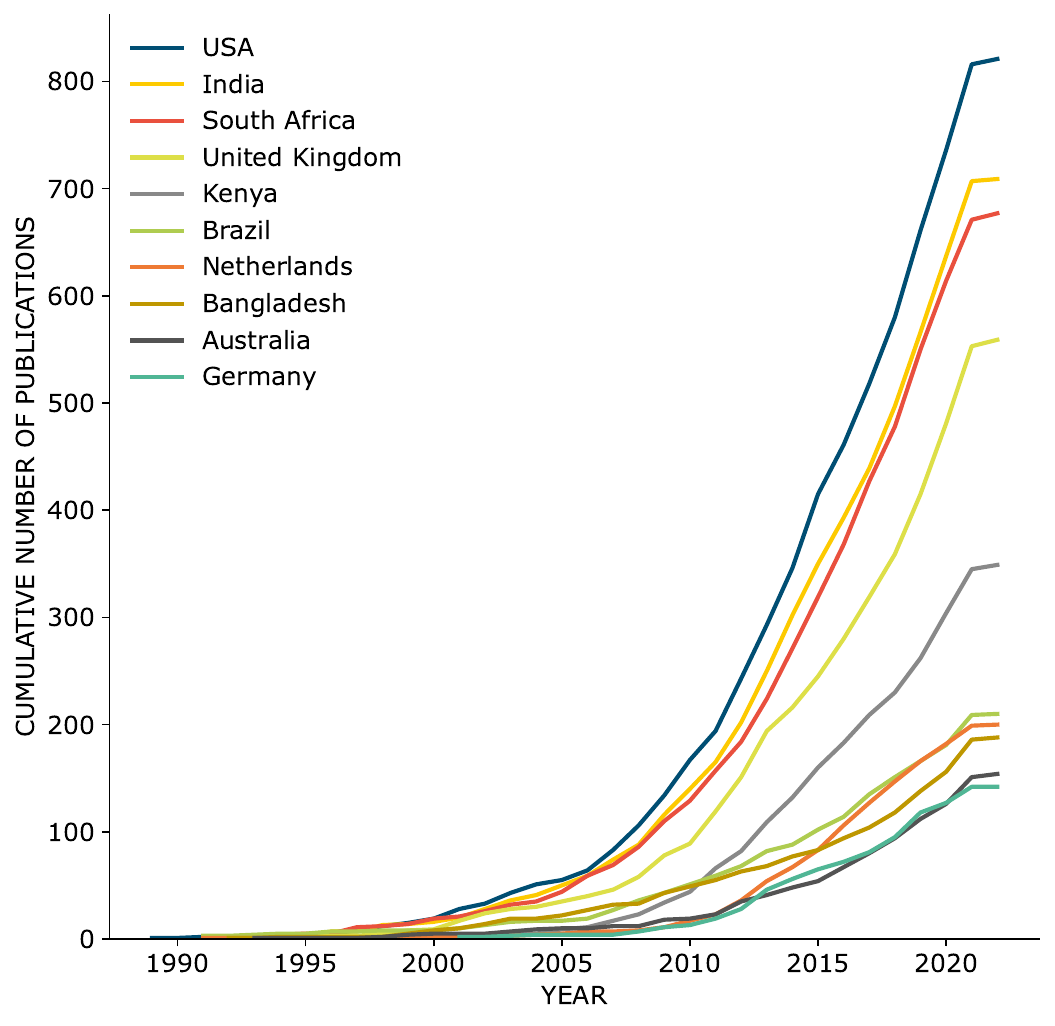}
\caption{Temporal evolution of researching institution countries}
\label{fig:temporalResearchingCountries}
\end{figure}

\begin{figure}[h!]
\centering
\includegraphics[width=1\textwidth]{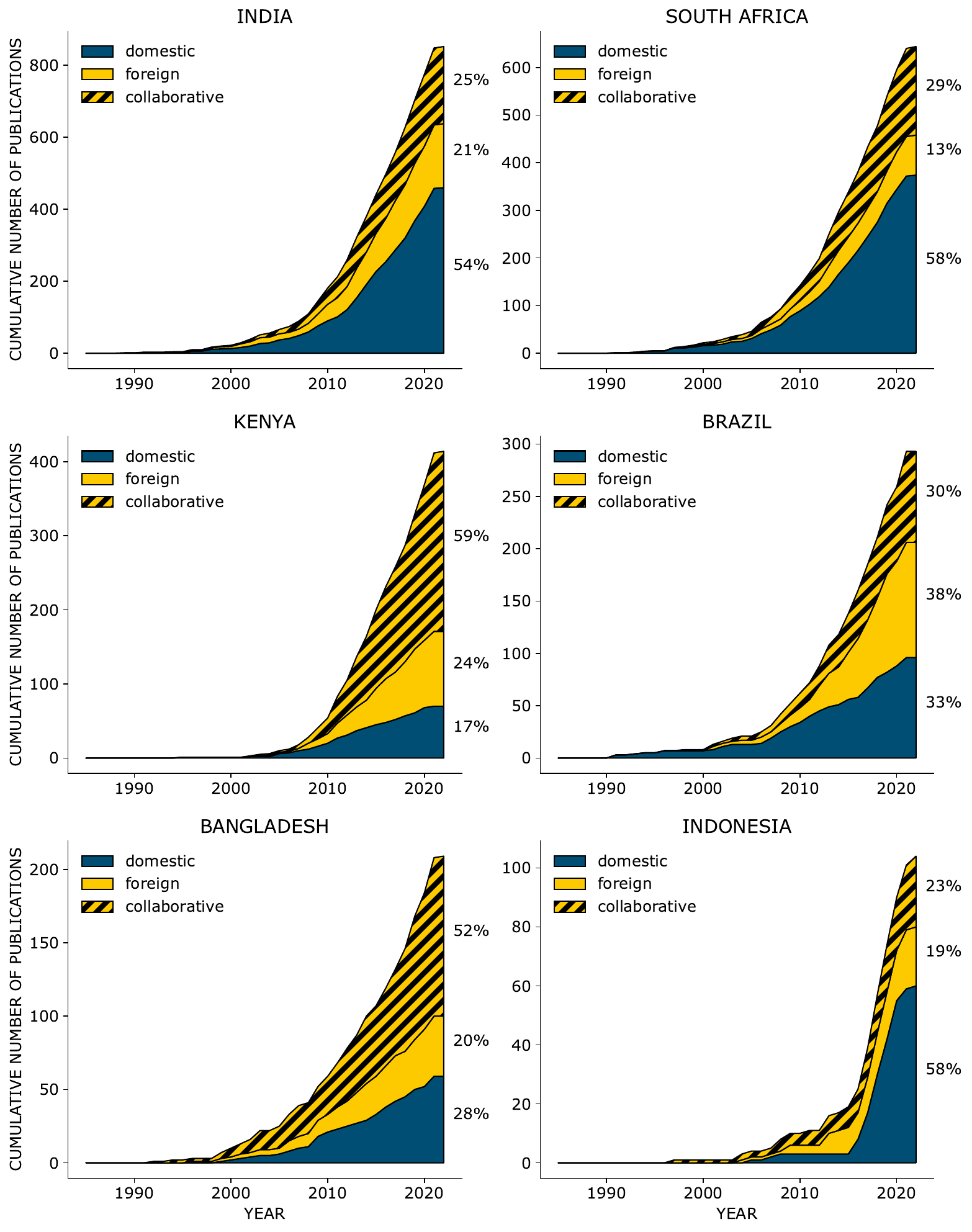}
\caption{Temporal development of studies for India, South Africa, Kenya, Brazil, Bangladesh and Indonesia and their division in \textit{domestic}, \textit{foreign} and \textit{collaborative}. The percentages doesn't necessary have to be equal to the percentages in Figure~\ref{fig:reasearched_bar}, since there are also publications without information on the publication year.}
\label{fig:mixed_foreign_native_over_years}
\end{figure}

\FloatBarrier

\bibliographystyle{unsrt}
\bibliography{references.bib}

\end{document}